%% file: cr_clean.tex
\documentclass[useAMS,usenatbib]{mn2e}
\usepackage{graphicx}
\usepackage{color}
\usepackage{amssymb}
\usepackage{amsmath}
\usepackage{upgreek}
\usepackage{subfigure}
\usepackage{rotating}
\usepackage{pdflscape}
\usepackage{bm} 
\usepackage{mathtools}

\usepackage{comment}
\excludecomment{details}

\title[The ionising effect of stellar CRs in PPDs]{ The ionising effect of low energy cosmic rays from a class II object on its protoplanetary disk}
\author[Rodgers-Lee, Taylor, Ray, Downes]{D. Rodgers-Lee$^{1,2,3}$\thanks{E-mail:
donna@cp.dias.ie}, A. M. Taylor$^{1}$, T. P. Ray$^{1,2}$, T. P. Downes$^{4,5}$ \\ 
$^{1}$ School of Cosmic Physics, Dublin Institute for Advanced Studies, 31 Fitzwilliam Place, Dublin 2, Ireland \\
$^{2}$ School of Physics, Trinity College Dublin, Dublin 2, Ireland \\
$^{3}$ Centre for Astrophyics Research, School of Physics, Astronomy and Mathematics, University of Hertfordshire,\\ College Lane, Hatfield AL10 9AB, UK\\
$^{4}$ School of Mathematical Sciences, Dublin City University, Glasnevin, Dublin 9, Ireland\\
$^{5}$ National Centre for Plasma Science and Technology, Dublin City University, Glasnevin, Dublin 9, Ireland }
\begin{document}
\date{Accepted 2017 July 24. Received 2017 July 06; in original form 2016 December 01}
\pagerange{\pageref{firstpage}--\pageref{lastpage}} \pubyear{2017}
\maketitle

\label{firstpage}

\input{abstract}


\input{intro}

\input{form}

\input{results}

\input{conclusions}

\input{acknowledgements}

\appendix
\input{appendix}

\newcommand\aj{AJ} 
\newcommand\actaa{AcA} 
\newcommand\araa{ARA\&A} 
\newcommand\apj{ApJ} 
\newcommand\apjl{ApJ} 
\newcommand\apjs{ApJS} 
\newcommand\aap{A\&A} 
\newcommand\aapr{A\&A~Rev.} 
\newcommand\aaps{A\&AS} 
\newcommand\mnras{MNRAS} 
\newcommand\pasa{PASA} 
\newcommand\pasp{PASP} 
\newcommand\pasj{PASJ} 
\newcommand\solphys{Sol.~Phys.} 
\newcommand\nat{Nature} 
\newcommand\bain{Bulletin of the Astronomical Institutes of the Netherlands}
\newcommand\memsai{Mem. Societa Astronomica Italiana}
\newcommand\apss{Ap\&SS} 
\newcommand\qjras{QJRAS} 
\newcommand\pof{Physics of Fluids}
\newcommand\grl{Geophysical Research Letters}
\newcommand\ssr{Space Science Review}
\bibliographystyle{mn2e}
\bibliography{donnabib}

\label{lastpage}

\end{document}

%% file: abstract.tex
\begin{abstract}

We investigate the ionising effect of low energy cosmic rays (CRs) from a young star on its protoplanetary disk (PPD).  We consider specifically the effect of $\sim3$\,GeV protons injected at the inner edge of the PPD. An increase in the ionisation fraction as a result of these CRs could allow the magnetorotational instability to operate in otherwise magnetically dead regions of the disk. For the typical values assumed we find an ionisation rate of $\zeta_\mathrm{CR} \sim 10^{-17}\mathrm{s^{-1}}$ at $1\,$au.

The transport equation is solved by treating the propagation of the CRs as diffusive. We find for increasing diffusion coefficients the CRs penetrate further in the PPD, while varying the mass density profile of the disk is found to have little effect. We investigate the effect of an energy spectrum of CRs. The influence of a disk wind is examined by including an advective term.  For advective wind speeds between $1-100\mathrm{km\,s^{-1}}$ diffusion dominates at all radii considered here (out to 10\,au) for reasonable diffusion coefficients.

Overall, we find that low energy CRs can significantly ionise the midplane of PPDs out to $\sim\,$1\,au. By increasing the luminosity or energy of the CRs, within plausible limits, their radial influence could increase to $\sim$2\,au at the midplane but it remains challenging to significantly ionise the midplane further out. 

\end{abstract}

%% file: intro.tex
\section{Introduction}
\label{sec:intro}

Low-mass star formation is an area of great interest as it allows us to study the physical conditions under which the solar system may have formed. Young stellar objects (YSOs) were initially separated into three main categories \citep{adams_1987,lada_1987}. Class I objects represent one of the earliest stages of low-mass star formation where the protostar is occluded by an envelope of infalling gas and dust. Class II objects, also known as classical T Tauri stars (CTTSs), consist of a young central star surrounded by a circumstellar disk and are the focus of this paper. At this stage the envelope has largely dissipated. The circumstellar disks surrounding CTTSs are thought to be the birthplace of planets and therefore are also generally known as protoplanetary disks (PPDs). Class III objects represent a later stage when the disk has mainly been accreted onto the central star or dispersed. Further classes were subsequently added, such as Class 0s which are thought to be younger than Class Is \citep{andre_1993} and represents the stage when most of the mass of the star has yet to be accreted.

CTTSs are observed to have mass accretion rates of the order of $10^{-8} M_\odot \mathrm{yr}^{-1}$ \citep[][for instance]{sicilia-aguilar_2004}. The mechanism which allows material to accrete from the PPD onto the central star still remains unclear, as the material must first lose angular momentum if it is to move radially inwards. If accretion is driven by a magnetically mediated process there are two main options: the magnetorotational instability \citep[MRI,][]{balbus_1991,hawley_1991} or magnetocentrifugally launched winds \citep[MCWs,][]{blandford_1982}. For these mechanisms the ionisation fraction in PPDs is of crucial importance as it dictates the strength of the coupling between the magnetic field and the material in the disk. The strength of this coupling in turn dictates the efficacy of the MRI and MCWs. Regions of the PPD where the magnetic field is not coupled to the neutral species are known as ``dead-zones", as proposed by \citet{gammie_1996}. The weak ionisation of large areas of PPDs leads to the necessity of including non-ideal magnetohydrodynamic (MHD) effects in numerical simulations of the MRI, the strength of which are dictated by the ionisation fraction of the PPD \citep{wardle_2012}.

Until recently the sources of ionisation considered have been X-rays \citep{igea_1999} and FUV radiation \citep{perez-becker_2011} from the CTTS itself, Galactic cosmic rays \citep[GCRs,][]{umebayashi_1981} and radioactivity \citep{umebayashi_2009}. These sources of ionisation dominate in different regions of the disk: X-rays from the young star itself are attenuated quickly and thus are only important for the surface of the disk. Ionisation by radioactivity leads to an extremely low ionisation rate. GCRs with a range of energies were, until recently, thought to be important in the outer regions of the disk and could penetrate deep into the disk towards the midplane there.

As is well known, the solar wind suppresses low energy (up to $\sim$GeV) GCRs from the solar system, out to $\sim$100\,au \citep{webber_1998,webber_2013}. These low energy GCRs have been assumed as a source of ionisation for PPDs, particularly important for the outer regions of the disk. \citet{cleeves-2013} showed however that, analogous to the solar system, these particles would similarly be excluded by the `T-Tauriosphere', the so-called heliosphere of a CTTS. Winds from CTTSs are thought to be stronger than the solar wind due to increased levels of magnetic activity \citep{feigelson_1999}, meaning low energy CRs could be suppressed even further out to $\sim10^4$\,au \citep{cleeves-2013}.

 Currently our understanding of where, and to what extent, the MRI operates in PPDs relies critically on the assumed ionisation fraction. If low energy GCRs are suppressed, but cosmic rays (CRs) from the \textit{young star} instead are included, the ionisation fraction will vary differently as a function of radius and height above the disk. This has implications for non-ideal MHD simulations of PPDs.

The motivation for investigating the ionising effect of CRs from a CTTS comes from a number of observational clues suggesting that YSOs and the environments surrounding them are capable of accelerating particles up to at least $\sim$GeV. For example, an unexpectedly high ionisation rate was inferred from chemical reactions surrounding a Class 0 object \citep{ceccarelli_2014} and this was explained by CRs produced via diffusive shock acceleration at either the protostellar surface or in jet shocks \citep{padovani_2015,padovani_2016}. There is also evidence for non-thermal emission from the bow shock of a knot in the jet of a Class II source \citep{ainsworth_2014} indicating the presence of $\sim$GeV CRs associated with Class II systems.

In this paper, we consider the ionising effect of low energy CRs (produced by the CTTS itself) on its PPD. This was previously studied by \citet{turner_2009}, and more recently by \citet{rab_2017}, but the difference in our approach will be discussed in Section\,\ref{sec:form}. Section\,\ref{sec:form} also introduces the numerical method, along with the initial conditions used and a discussion of the parameters involved. We present our results in Section\,\ref{sec:results} and discuss our results in comparison to observations and in particular the known stellar X-ray properties of YSOs in Section\,\ref{sec:discussion}. Finally, we form our conclusions in Section\,\ref{sec:conclusions}.

%% file: form.tex
\section{Formulation}
\label{sec:form}

\begin{figure}
\centering
 \includegraphics[width=0.45\textwidth]{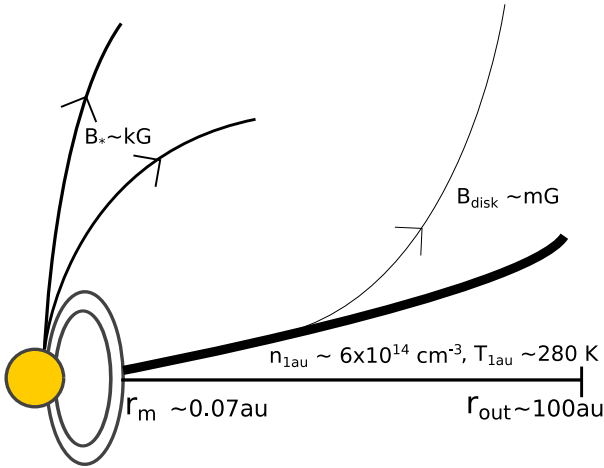}
  \caption{Schematic of the system we consider with typical values for the mass density, temperature and magnetic field strength for the protoplanetary disk and the central young star. Axial symmetry is assumed about the axis of rotation and mirror symmetry about the midplane of the disk.}
\label{fig:disk-schematic}
\end{figure}

To investigate the ionising effect of low energy CRs, originating from the young star itself, in PPDs we model their propagation as diffusive \citep[as opposed to rectilinear transport as in][which results in a geometric dilution factor of $r^{-2}$]{turner_2009,rab_2017} due to the turbulent magnetic fields thought to exist at the inner edge of PPDs. The magnetic field is thought to be turbulent here as a result of MRI driven turbulence which can develop due to thermal ionisation \citep{gammie_1996} of the disk. Fig.\,\ref{fig:disk-schematic} gives some of the typical physical quantities for the system considered. 

Measuring the magnetic field strength or its orientation in PPDs is challenging but observations of molecular clouds indicate $\mu$G to mG magnetic fields are present \citep{crutcher_2012}. By examining chondrites in the solar system it is possible to estimate the strength of the magnetic field in which they formed, probably at distances of $\sim$1\,au in the Sun's PPD, to be $0.1-10$G \citep{levy_1978}.

Therefore, by treating the propagation of the CRs as diffusive we have tried to include the effect that turbulent magnetic fields would have. In the presence of turbulent magnetic fields CRs are thought to scatter off anisotropies in the field which is the basis of quasi-linear theory \citep{jokipii-1966,schlickeiser_1989}. This random walking of the CRs can be modelled as diffusive transport as described by Eq.\,\ref{eq:ncr} in Section\,\ref{subsec:diffusion}.  Treating the propagation as diffusive, in the absence of energy losses in the system, would result in a geometric dilution factor of $r^{-1}$. This potentially makes CRs much more important than X-rays as a source of ionisation further away from the star as their number density does not drop as sharply with increasing radius due to the difference in the geometric dilution factor. We only consider the propagation of protons since radiative cooling for electrons inhibits their acceleration to higher energies making them much less important for the overall ionisation rate. 

The X-ray luminosity of CTTSs is significantly larger than that of the Sun. Therefore, we make the assumption that CTTSs are able to accelerate particles up to $\sim$GeV energies, in the same way that the Sun is an effective MeV accelerator \citep[][for instance]{mewaldt_2005}. This is discussed further in Section\,\ref{subsubsec:cr-spectrum}. 

In this paper, we focus predominantly on the ionising effect that $\sim$3\,GeV CRs would have on the PPD and do not in general consider a spectrum of CRs. We do also briefly consider the effect that higher energy CRs ($<300$\,GeV) would have, if present in the system. Section \ref{subsubsec:cr-spectrum} describes how we construct a spectrum for the CRs in this case.

As mentioned above, in the presence of a turbulent magnetic field the propagation of CRs may be treated as a diffusive process. Not all regions of the disk are thought to be turbulent though. For instance, a large-scale ordered vertical magnetic field threading the disk may allow a MHD wind to be driven high up in the disk. This could advect the CRs away from the disk. To investigate this effect, we also include an advective flow to mimic the presence of a MHD wind. 

\subsection{Transport equation}
\label{subsec:diffusion}
We solve the transport equation for the CRs which treats their propagation as diffusive and is given by
\begin{equation}
\frac{\partial n_\mathrm{CR}}{\partial t} = \bm{\nabla}\cdot(D\bm{\nabla} n_\mathrm{CR}) - \frac{n_\mathrm{CR}}{\tau} - \bm{v}\cdot\bm{\nabla} n_\mathrm{CR} + Q
\label{eq:ncr}
\end{equation}
\noindent where $n_\mathrm{CR}$ is the number of cosmic rays, $D(r,z)$ is the diffusion coefficient, $1/\tau(r,z)$ is the CR loss rate which is related to the mass density of the PPD, $\bm{v}$ is the advection speed (this term is only relevant when an advective wind is considered) and $Q(r,z)$ is the number of CRs injected per unit time per unit volume which is normalised to one.  Each of these quantities are discussed further in Section\,\ref{subsec:params}. The corresponding difference equation for Eq.\,\ref{eq:ncr} is given in Appendix\,\ref{appendix:dif-eq}. We solve Eq.\,\ref{eq:ncr} in cylindrical coordinates assuming axial symmetry.

Once the simulation has reached a steady-state throughout the computational domain the losses in the system are in equilibrium with the injection rate. 
\subsubsection{Numerical method}
We advance Eq.\,\ref{eq:ncr} in time using a forward in time, centred in space differencing scheme for the diffusive term. To advance the advective term we use the Lax-Wendroff scheme \citep{lax_1960}. The overall scheme is second order in space and first order in time. A numerical convergence test for the scheme is presented in Appendix\,\ref{appendix:res}.

The outer $r$ and upper $z$ boundaries are absorptive allowing the CRs to diffuse/advect out of the system. We also assume mirror symmetry about the midplane of the PPD so only the upper half of the PPD is simulated, from the midplane up to some height $z$ above the midplane of the disk (the values of $z$ used are given in Table\,\ref{table:sims}). Therefore, the inner $r$ and the lower $z$ boundaries are reflective due to the assumption of axial symmetry about the axis of rotation and the midplane of the PPD, see Fig.\,\ref{fig:disk-schematic}.

\subsubsection{Temporal convergence test}
Once the simulations have reached a steady state then Eq.\,\ref{eq:ncr} becomes
\begin{equation}
\frac{n_\mathrm{CR}}{\tau} + \bm{v}\cdot\bm{\nabla} n_\mathrm{CR} - \bm{\nabla}\cdot(D\bm{\nabla} n_\mathrm{CR}) = Q 
\label{eq:ncr-ss}
\end{equation}
This states that the amount of CRs injected into the computational domain is balanced by the amount of CRs lost to ionisation (the sink term), along with diffusion and advection. To check that the simulations are temporally converged we examine the total number of CRs in the computational domain as a function of time. Once this quantity converges the simulations have reached a steady state. If the simulations reach a steady state via the sink term then the total number of CRs lost per time step is equivalent to the total number of CRs remaining in the computational domain (per time step).

\subsubsection{Low density limit}
\label{appendix:low-density}
\begin{figure}
\centering
 \includegraphics[width=0.5\textwidth]{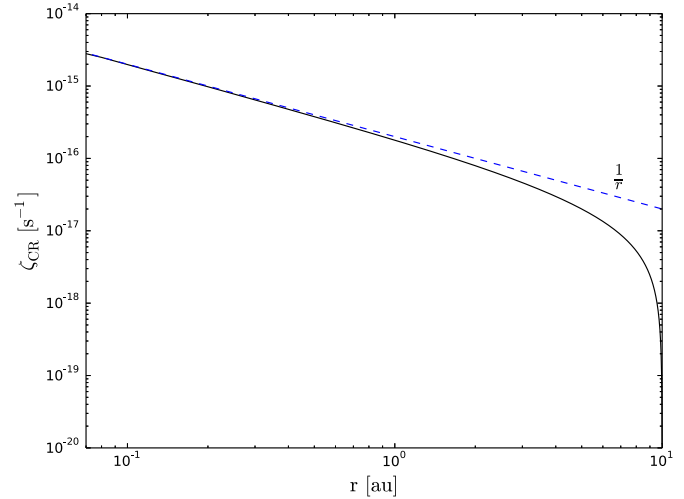}
  \caption{Ionisation rate plotted as a function of radius for the constant low density test case (solid black line) where the analytic result for diffusion in the absence of a sink term is recovered, giving the $r^{-1}$ profile (dashed blue line).}
\label{fig:low_density_test}
\end{figure}

In the absence of a sink term or when the sink term is very small, corresponding physically to a low mass density in the disk, then the radial profile of the CRs should develop an $r^{-1}$ dependence. By setting the density to be a constant, $\rho=\rho_0\times 10^{-8}$, in the computational domain for simplicity results in the $r^{-1}$ dependence as expected, as shown in Fig.\,\ref{fig:low_density_test}. Here, $\rho_0 = 2.33 \times 10^{-9}\mathrm{g\,cm^{-3}}$ \citep[from][]{salmeron_2003}. Note that the drop-off at 10\,au is due to the interaction of the CRs with the boundaries.

\subsubsection{Effect of computational domain size}
\label{appendix:comp_dom}
All the simulation results presented in Section\,\ref{sec:results} use a 10\,au box size. To test the effect of the size of the box on the results we ran a 20\,au run, d030-adv0-p1-res01-rad20, details of which are given in Table\,\ref{table:sims}. We found that the simulations varied $\lesssim 10\%$ at 8\,au and $\lesssim 20\%$ at 9\,au. At smaller radii there are $\lesssim 1\%$ differences. The flux of CRs reaching the outer regions is also too low for us to consider as being important in terms of an ionisation rate. We conclude that the boundary conditions are having minimal effect on the simulation results and the conclusions drawn from them.

\subsection{Parameters}
\label{subsec:params}
\subsubsection{Number density of cosmic rays}
\label{subsubsec:ncr} 
One of the quantities that we must estimate for these simulations is the number density of CRs. To do this we assume that X-ray luminosity ($L^\mathrm{X}$) and CR luminosity ($L^\mathrm{CR}$) are correlated. We consider the ratio of $L^\mathrm{X}_\odot$ and $L_\odot^\mathrm{CR}$ for the Sun and then scale this to a CTTS using the X-ray luminosity of a typical young star ($L_{*}^\mathrm{X}$). We estimate $L_\odot^\mathrm{CR}$ by assuming that $L_\odot^\mathrm{CR} \sim L^\mathrm{SW}_\odot$, the solar wind luminosity. Particles from the solar wind are known to be accelerated at the heliospheric termination shock \citep[][for instance]{decker_2008}, whereas in our case the protons are assumed to be accelerated much closer to the star.

For the Sun, $L^\mathrm{X}_\odot \sim 5 \times 10^{27} \mathrm{erg\,s^{-1}}$ \citep{peres-2000} at solar maximum. At solar minimum this value is an order of magnitude smaller but during a very large solar flare $L^\mathrm{X}_\odot \sim 2 \times 10^{28} \mathrm{erg\,s^{-1}}$. We take the solar wind luminosity to be $L^\mathrm{SW}_\odot \sim 1 \times 10^{27} \mathrm{erg\,s^{-1}}$ \citep{drury-2012}. We adopt a value of $L_{*}^\mathrm{X}\sim 1 \times 10^{29} \mathrm{erg\,s^{-1}}$ \citep{feigelson_1999} though this could reasonably be increased by two orders of magnitude.

Finally, we arrive at an estimate for the CR luminosity which is given below
\begin{equation}
L_{*}^\mathrm{CR} \sim L^\mathrm{CR}_\odot \left(\frac{L^\mathrm{X}_*}{L^\mathrm{X}_\odot}\right) \sim 1 \times 10^{28} \mathrm{erg\,s^{-1}}
\label{eq:ncr-lum}
\end{equation}

\citet{ainsworth_2014} estimated the luminosity from low energy electrons for the bow shock of DG Tau (a class II source) to be $L_\mathrm{e} \sim 3 \times 10^{29}\mathrm{erg\,s^{-1}}$, an order and a half magnitude larger than our estimate above. From \citet{ainsworth_2014} the ratio of the total proton energy density to electron energy density, $k$, was taken to be 40 corresponding to a non-relativistic strong shock regime. Therefore we can estimate the luminosity from low energy protons, $L_\mathrm{p}$, for the bow shock of DG Tau  as $L_\mathrm{p}=40L_\mathrm{e} \sim 1 \times 10^{31}\mathrm{erg\,s^{-1}}$. This luminosity is appreciably larger than the value we have assumed above and indicates the possibility of reasonably increasing our assumed value above by at least an order of magnitude. Note however, that the acceleration region for the non-thermal emission detected in the bow shock in the jet of DG Tau is significantly different to the location considered in this paper but does indicate the presence of low energy CRs in Class II systems.

\subsubsection{Diffusion coefficient}
The diffusion coefficient determines the extent to which the CRs can propagate through the disk and is energy dependent. The diffusion coefficient, expressed in units of the CR's velocity (which for relavistic CRs is $\sim c$), can be estimated from quasi-linear theory \citep{jokipii-1966,schlickeiser_1989} as 

\begin{equation}
\frac{D}{c} = \beta\eta(p)r_\mathrm{L}
\label{eq:diff-coeff}
\end{equation}
\noindent where $r_\mathrm{L}$ and $p$ are the Larmor radius and momentum of the CR respectively. $\beta$ is the velocity of the CR given as a fraction of the speed of light. $\eta(p)$ describes the level of turbulence in the magnetic field and is given by
\begin{equation}
\eta(p) = \eta_0\left( \frac{p}{p_0}\right)^{1-\gamma}
\label{eq:dif-norm}
\end{equation}

 where as $\eta \rightarrow 1$, Bohm diffusion is recovered and a particle with  momentum $p$ scatters once per Larmor radius. This sets the lower limit on the diffusion coefficient (in units of the speed of light) as the Larmor radius of the proton. For a proton with momentum $3\,\mathrm{GeV}/c$ in a mG magnetic field, the Larmor radius $r_\mathrm{L} \sim 3 \times 10^{-4}$\,au. We explore the effect of a range of values for $D$ from $3 - 380r_\mathrm{L}$ by varying $\eta_0$ for 3\,GeV CRs, details of which are shown in Table\,\ref{table:sims}. The parameter $\gamma$, in Eq.\,\ref{eq:dif-norm}, reflects the underlying physical mechanism for the turbulence in the magnetic field where for a Kolmogorov (hydrodynamic driven) turbulence spectrum $\gamma=5/3$ or for a Kraichnan type spectrum \citep[MHD driven,][]{kraichnan_1965} $\gamma=3/2$. 

\subsubsection{Cosmic ray loss rate}
\label{subsubsec:sink-term}
The CR loss rate ($1/\tau(r,z)$ from Eq.\,\ref{eq:ncr}) is the rate at which the CRs lose their energy. The loss rate depends on the energy of the CR and on the type and density of the material that it is travelling through. 

We express the loss rate as a function of grammage, $\chi$, where $\chi=\rho x$ with $\rho$ representing the mass density of the plasma (in this case the disk) and $x$ is depth. The loss rate as a function of grammage for a $\sim$GeV CR can be expressed as
\begin{equation}
\frac{1}{\tau_\mathrm{GeV}} = \frac{1}{E_\mathrm{GeV}}\frac{dE_\mathrm{GeV}}{dt}=\frac{1}{E_\mathrm{GeV}}c\rho\left( \frac{dE_\mathrm{GeV}}{d\chi}\right)_p
\label{eq:loss-rate}
\end{equation}
where the mean energy loss rate for a $\sim$GeV proton in $\mathrm{H_2}$ is $\left( \dfrac{dE_\mathrm{GeV}}{d\chi}\right)_p\sim 2\times 10^{-3} \mathrm{GeV \,g^{-1}\,cm^{2}}$ \citep{longair_2011}. At 1\,au, the loss rate is $1/\tau_\mathrm{1au}=0.14\,\mathrm{s^{-1}}$ for a $\sim$GeV CR. The mass density profile of the PPD used is
\begin{equation}
\rho = \rho_0e^{-r_\mathrm{in}^2/r^2}\left(\frac{r_0}{r} \right)^p e^{-z^2/2H^2} + \rho_\mathrm{ISM}
\label{eq:rho}
\end{equation}
\noindent where $\rho_0 = 2.33 \times 10^{-9}\mathrm{g\,cm^{-3}}$ \citep[from][]{salmeron_2003}, $r_\mathrm{in}=0.07\,$au is the inner edge of the disk (taken to be the magnetospheric radius, $r_\mathrm{m}$, described in Section\,\ref{subsubsec:cr-injection-site}), $r_0=1\,$au, $p=1.0$, $\rho_\mathrm{ISM}=3.89\times10^{-24} \mathrm{g\,cm^{-3}}$ is the mass density of the interstellar medium and $H$ is the pressure scale height of the disk defined as
\begin{equation}
H = \frac{c_s}{\Omega}
\label{eq:psh}
\end{equation}
where $\Omega=\sqrt{GM/r}$ is the Keplerian frequency, $G$ is the gravitational constant and $c_s$ is the sound speed which depends on the temperature, $T$, of the disk in the following way
\begin{equation}
c_s = \sqrt{\frac{k_\mathrm{B}T(r)}{\mu m_\mathrm{H}}}
\end{equation}
where $\mu=2.33$, representing a gas of hydrogen and helium and
\begin{equation}
T(r) = T_0\left( \frac{r_0}{r}\right)^q
\end{equation}
The parameter $q$ determines the temperature profile of the disk and usually $0.5<q<0.75$ \citep{chiang_1997,andrews_2005} with $T_0=280$\,K at 1\,au.

\subsubsection{CR injection site}
\label{subsubsec:cr-injection-site}
{We assume a source term of $\sim$GeV protons at the inner edge of the accretion disk where the stellar magnetic field, $B_*$, truncates the disk.} The magnetospheric radius ($r_\mathrm{m},$ also known as the Alfv\'en radius) occurs at the distance where the magnetic pressure from the young star and the ram pressure of the accretion flow balance. Following \citet{hartmann_2009}, and expressing it in a similar form to \citet{manara_2014}, $r_\mathrm{m}$ can be estimated as
\begin{equation}
r_\mathrm{m} \sim \left( \frac{3B^2_*R_*^6}{2\dot{M}_{\mathrm{acc}}\sqrt{GM_*}}\right)^{2/7} \sim 0.07\,\mathrm{au}
\end{equation}
\noindent We have taken $B_* \sim 1$kG \citep{johnstone_2014}, $R_*=2.5R_\odot$, $\dot{M}_{\mathrm{acc}} \sim 5 \times 10^{-8}M_\odot \mathrm{yr^{-1}}$ and $M_* = M_\odot$ which are typical values for a CTTS.

\subsubsection{Advection term}
\label{subsubsec:adv}
The ability to treat the propagation of CRs as a diffusive process, as we have in this paper, relies on a turbulent magnetic field. At some height above the disk though an MHD wind may be present, with ordered magnetic field lines, where the CRs may be advected. To simulate the effect of ordered magnetic fields from a MHD disk wind on the propagation of the CRs we include an advection term in Eq.\,\ref{eq:ncr}. Theoretically, MHD winds may be launched centrifugally when the angle between the magnetic field and the disk is $<60^\circ$ \citep{blandford_1982}. Therefore, we consider a wind at $45^{\circ}$ from the disk for simplicity. The vertical launching point of an MHD wind remains a topic of great interest. In this work, the advection term only operates at $z>3H$ above the disk since global non-ideal MHD simulations have found the wind to be launched approximately at this height \citep{gressel_2015}. Note, this launching height is only an estimate and depends sensitively on the strength of non-ideal effects which in turn depend on the ionisation fraction of the disk. Finally, we also vary the velocity of the wind to investigate the effect.

In the derivation of Eq.\,\ref{eq:ncr} the velocity field is assumed to be divergence free and so the velocity varies in the radial direction in the following way
\begin{equation}
v_r = v_0\left(\frac{r_0}{r}\right)
\label{eq:vr}
\end{equation}
where $v_0$ is the velocity at $r_0=1\,$au. At the same time, the vertical component of the wind is assumed to be constant. We vary the velocity between $1-10^4\,\mathrm{km\,s^{-1}}$, where $1-10^2\,\mathrm{km\,s^{-1}}$ represents the range of velocities expected from a slow molecular outflow to a highly collimated jet \citep{prpl-2014}. We also include an unrealistically high velocity of $10^4\,\mathrm{km\,s^{-1}}$ for illustrative purposes. 

Note, that the advection term included does not represent the advection associated with the inward accretion flow of the disk itself which we do not include as it is negligible in this case. To show this, we can estimate the velocity of the accretion flow, $v_\mathrm{r}$, for a typical observed mass accretion rate of $\dot{M} \sim 10^{-8} M_\odot \mathrm{yr}^{-1}$ for a CTTS. The mass accretion rate, at a particular radius $r$ and between a height $s$ above and below the disk midplane, can be expressed as
\begin{equation}
\dot{M} = 2\pi \int \limits_{-s}^{+s}\rho v_\mathrm{r} rdz
\label{eq:mdot}
\end{equation}
assuming axial symmetry for the disk. This can be rearranged to obtain an expression for $v_\mathrm{r}$ where we assume that $v_\mathrm{r}$ does not vary vertically. For $r=1\,$au and $s=\sqrt{2}H$, using the mass density profile given in Eq.\,\ref{eq:rho}, we find that $v_\mathrm{r}=3\times 10^{-5}\,\mathrm{km\,s^{-1}}$ which in comparison to the advective wind speeds considered ($1-100\,\mathrm{km\,s^{-1}}$) can be seen to be negligible. Section\,\ref{subsec:advection-results} shows how the advective timescales, even for the advection wind speeds, are much longer than the diffusive timescales considered in this work. Ultimately, this means that neither the advection wind speeds nor the accretion flow are important for the parameters investigated in this work.

\subsection{Diagnostics}
\subsubsection{Ionisation rate}
\label{subsubsec:ionisation-rate}
The ionisation rate, $\zeta_\mathrm{CR}$, is a more useful quantity to examine in the context of this paper rather than the number density of CRs hence this section describes how we convert $n_\mathrm{CR}$ to $\zeta_\mathrm{CR}$. The ionisation rate can be used to calculate an ionisation fraction which in turn indicates whether the MRI is able to operate in the disk. The ionisation rate is overall more useful to present here because the ionisation fraction depends on the particular chemical network model used to calculate the recombination rates. We do estimate the ionisation fraction as well but with some very basic simplifications described in Section\,\ref{subsubsec:ionisation-fraction}.

The unmodulated GCR ionisation rate is calculated from Eq.\,23 of \citet{umebayashi_1981} as $\sim 1 \times 10^{-17}\mathrm{s^{-1}}$ which we can compare our results with. We adapt Eq.\,23 from \citet{umebayashi_1981} to consider CRs of a single energy ($\sim3$GeV). We can then express the ionisation rate per hydrogen molecule in terms of $n_\mathrm{CR}$ and the energy-loss rate per grammage for protons in $\mathrm{H}_2$ as
\begin{eqnarray}
\zeta_\mathrm{CR}(r,z) = \frac{1}{{\overline{ \mathcal{E}}}}\left(\frac{\rho}{n} \left(\frac{dE}{d\chi}\right)_p \right) \Omega \frac{c}{4\uppi} n_\mathrm{CR}(r,z) \nonumber \\
= \frac{2m_\mathrm{H}}{\overline{ \mathcal{E}}} \left(-\frac{dE}{d\chi}\right)_\mathrm{p} \Omega \frac{c}{4\uppi} n_\mathrm{CR}(r,z)
\label{eq:zeta}
\end{eqnarray}

\noindent where $\overline{ \mathcal{E}}$ is the average energy loss per ionisation of a hydrogen molecule including the contribution from secondary electrons, $n$ is the number density of the PPD, $2m_\mathrm{H}$ is the mass of $\mathrm{H}_2$ and $\Omega$ is the solid angle. The loss rate per grammage for protons is the same as is used in Eq.\,\ref{eq:loss-rate}. Assuming $\Omega=4\uppi$ and using Eq.\,21 of \citet{umebayashi_1981} for $\sim3$GeV protons, along with the loss rate per grammage we can simplify Eq.\,\ref{eq:zeta} so that it becomes 
\begin{equation}
\zeta_\mathrm{CR}(r,z) = 2.2 \times 10^{-18} \mathrm{s^{-1}}\left( \frac{n_\mathrm{CR}(r,z)}{4 \times 10^{-10} \mathrm{cm^{-3}}} \right)
\label{eq:ion-rate}
\end{equation}

\subsubsection{Ionisation fraction}
\label{subsubsec:ionisation-fraction}
To calculate an ionisation fraction, $x_\mathrm{e}$, the net sources of ionisation and recombination must be considered. Without using a detailed chemical network, an estimate of $x_\mathrm{e}$ can be obtained by making a number of assumptions. Using the simplified chemical network from \citet{fromang_2002} and assuming chemical equilibrium the ionisation fraction is given by
\begin{equation}
x_\mathrm{e}(r,z) = \sqrt{\frac{\zeta_\mathrm{CR}(r,z)}{\beta_\mathrm{x} n(r,z)}}
\label{eq:ionfrac}
\end{equation}
where $\beta_\mathrm{x}$ is the recombination rate coefficient and $n(r,z)$ is the number density of the plasma. Eq.\,\ref{eq:ionfrac} is the simplest expression for $x_\mathrm{e}$ and is only valid in the limit that metals are either dominant or entirely absent. The value of $\beta_\mathrm{x}$ depends sensitively on the composition of the plasma, in particular on the abundance of metal atoms in the gas. \citet{fromang_2002} consider three different ways for recombination to take place, each with a different corresponding recombination rate: electron capture via dissociative recombination with molecular ions ($\beta_\mathrm{d}$), radiative recombination with heavy-metal ions ($\beta_\mathrm{r}$) and charge transfer from molecular ions to metal atoms ($\beta_\mathrm{t}$). These possible recombination rate coefficients \citep[from][]{fromang_2002} are given by
\begin{eqnarray}
&\beta_\mathrm{r} &= 3 \times 10^{-11} T^{-1/2} \mathrm{cm^{3} s^{-1}} \nonumber \\
&\beta_\mathrm{d} &= 3 \times 10^{-6} T^{-1/2} \mathrm{cm^{3} s^{-1}} = 10^{5} \beta_\mathrm{r}\nonumber \\
&\beta_\mathrm{t} &= 3 \times 10^{-9} \mathrm{cm^{3} s^{-1}}
\end{eqnarray}
\noindent In Section\,\ref{sec:results} we will consider two extremes for calculating the ionisation fraction, when metals are present or absent in the gas. If metal ions are absent in the gas then the appropriate rate coefficient in Eq.\,\ref{eq:ionfrac} is $\beta_\mathrm{d}$, from \citet{fromang_2002} this case would correspond to the metal ions being locked in sedimented grains. On the other hand, if the metal ions dominate as charge carriers in the plasma then the appropriate rate coefficient is $\beta_\mathrm{r}$. The extent to which metal ions are locked in sedimented grains would depend on the specific disk considered, leading us to merely present the two extreme cases as indicative of the range of possible values for $x_\mathrm{e}$. Details of how $x_\mathrm{e}$, as given in Eq.\,\ref{eq:ionfrac}, is derived in the presence or absence of metals is given in \citet{fromang_2002}.

\subsubsection{CR energy spectrum}
\label{subsubsec:cr-spectrum}
While we concentrate predominantly on the effect of 3\,GeV CRs, we also consider the ionising effect of a range of CR energies in Section\,\ref{subsec:results_energy}. To construct a spectrum from CRs of different energies, or momenta, we use the following weighting relationship
\begin{equation}
\xi_\mathrm{i} = \left( \frac{p_\mathrm{i}}{p_0} \right)^{1-\alpha}e^{(-p_\mathrm{i}/p_\mathrm{max})} \hspace{1mm}
\label{eq:cr-weight}
\end{equation}
such that
\begin{equation}
\zeta_\mathrm{CR,i}(r,z) =  \frac{\xi_\mathrm{i}\zeta_\mathrm{CR,0}(r,z)}{\sum\limits_{i=1}^N  \xi_\mathrm{i} } 
\label{eq:xi}
\end{equation}

\noindent where $p_\mathrm{i}=0.1,...,300\,\mathrm{GeV}/c$ in Eq.\,\ref{eq:cr-weight} are the logarithmically spaced momentum bins for the CRs with $N=8$. $p_\mathrm{0}\,(=100\,\mathrm{MeV}/c)$ and $p_\mathrm{max}\,(=300\,\mathrm{GeV}/c$) are the minimum and maximum momenta of CRs considered, respectively. $\zeta_\mathrm{CR,i}(r,z)$ is the contribution to the total CR ionisation rate from CRs of a single momentum, $p_\mathrm{i}$, meaning also that $\zeta_\mathrm{CR,0}(r,z)$ is the contribution from CRs with momentum $p_0$.

The index $\alpha$ in Eq.\,\ref{eq:cr-weight} is related to the acceleration process, $\alpha=2$ is assumed here which represents a fiducial value for Fermi first order mechanisms. For comparison, \citet{rab_2017} used the same spectral slope as the work presented here \citep[derived from the MeV solar observations of ][]{mewaldt_2005,reedy_2012} but with a broken power law which transitions at $\sim 20$\,MeV to a significantly steeper slope. The spectral index of $\alpha=2$ up to MeV energies for low energy cosmic rays from the sun can be explained by diffusive shock acceleration or from magnetic reconnection or a combination of both \citep{reames_2013}. Note, the GCR spectrum possesses a significantly steeper (i.e. softer) spectral index than that adopted in our calculations, with a spectral index of $\sim$2.8 \citep[see the Alpha Magnetic Spectrometer results from][]{aguilar_2015}.

The $\sim 20$\,MeV break in the solar spectrum to a steeper power law has been seen to migrate to more than an order of magnitude higher energies during solar flare events \citep{ackermann_2014,ajello_2014}. This can be thought of as more representative of what we might expect the CR spectrum for a CTTS to be but extending up to GeV energies. The order of magnitude maximum energy we use is motivated by simple scaling arguments of the strength of the solar magnetic field and maximum energies of solar CRs and the corresponding magnetic field strength of the young star. We approximate the broken power law \citep[used in][but with the break at MeV energies]{rab_2017} as an exponential cut-off instead at the maximum energy considered.

%% file: results.tex
\section{Results}
\label{sec:results}
\begin{figure*}%
	\centering
    \subfigure[ ]{%
        \includegraphics[width=0.33\textwidth]{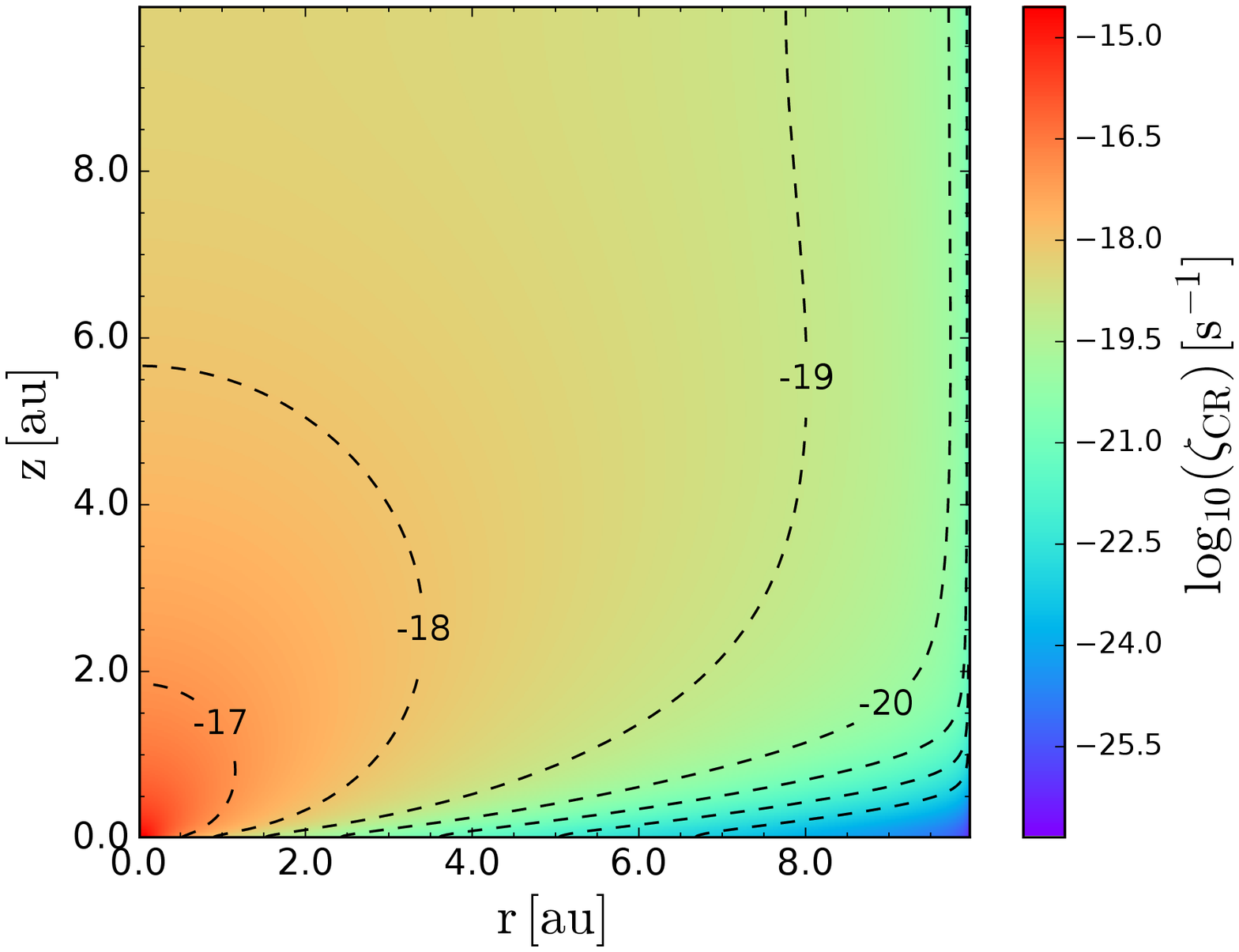}
       	\centering
\label{fig:ionrate_d02}}%
  	    ~
    \subfigure[ ]{%
        \includegraphics[width=0.33\textwidth]{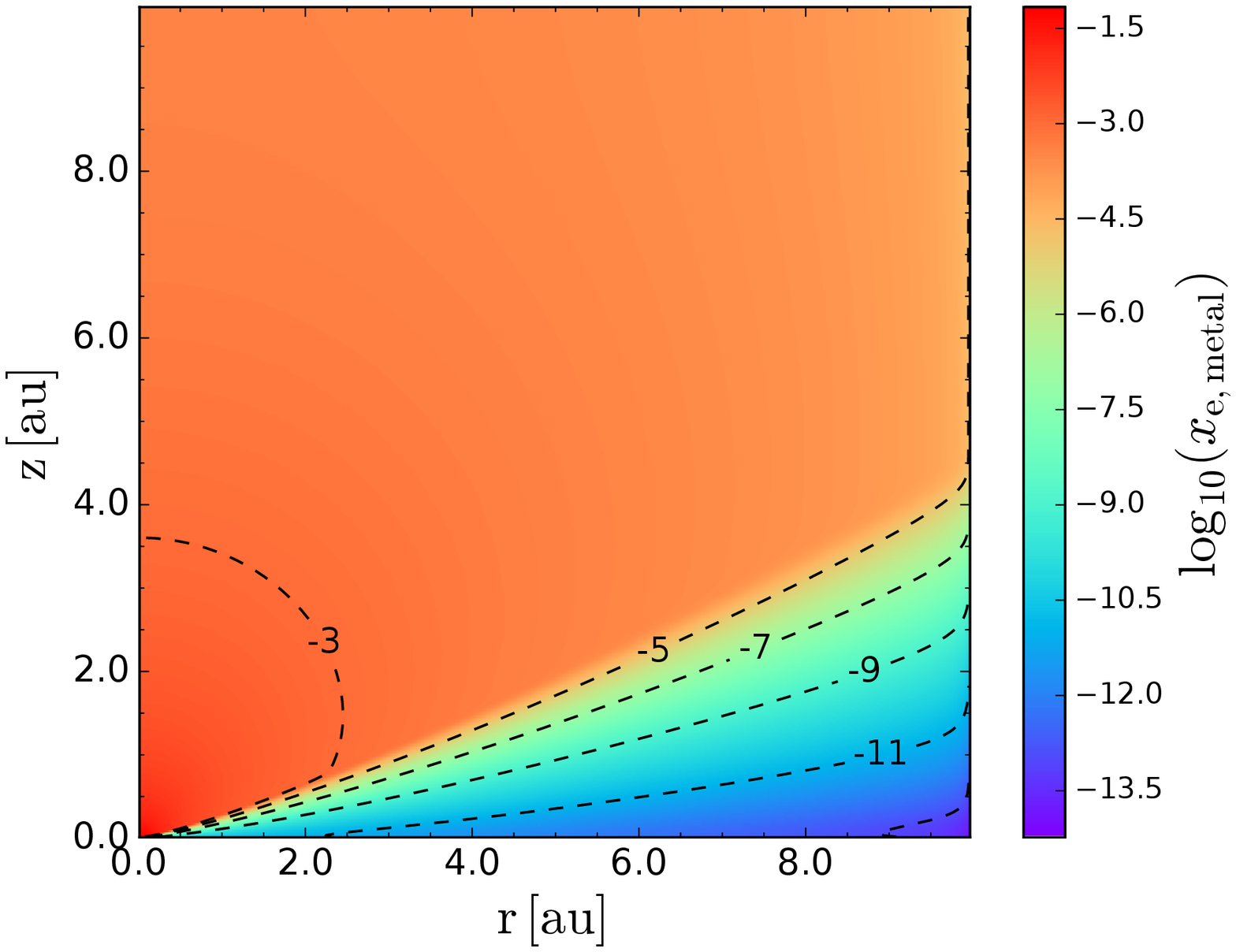}
	\centering
 \label{fig:ionfrac_metal}}%
   	    ~
    \subfigure[ ]{%
        \includegraphics[width=0.33\textwidth]{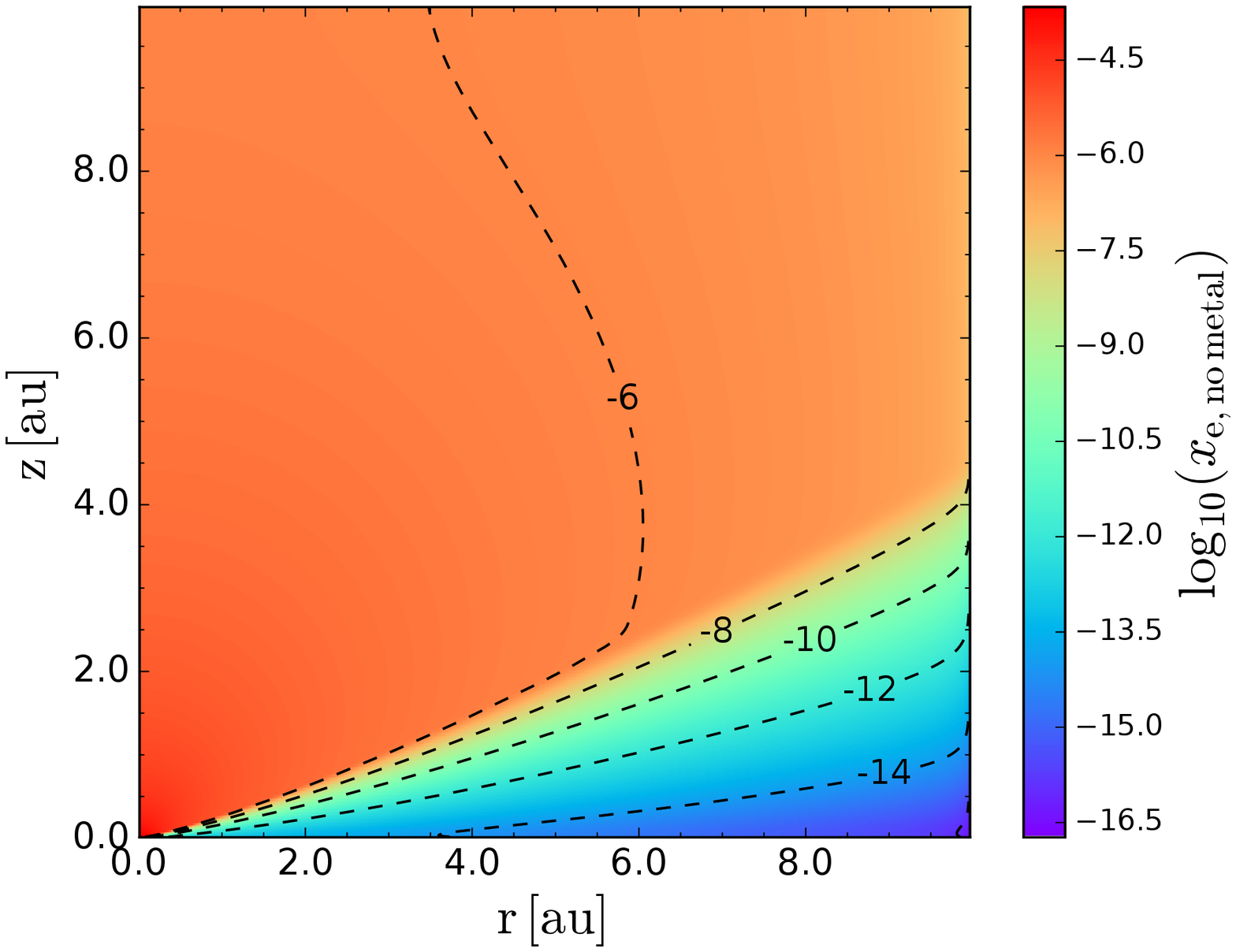}
	\centering
 \label{fig:ionfrac_nometal}}%
    \caption{(a) The ionisation rate, $\zeta_\mathrm{CR}$, for d030-adv0-p1-res01 is plotted as a function of radius and height with $D/c= 30r_\mathrm{L}$, (b) plots the ionisation fraction, $x_\mathrm{e,\,metal}$, when metal ions are present and (c) plots the ionisation fraction, $x_\mathrm{e,\,no\,metal}$, when metal ions are absent.} 
    \label{fig:fid_run}%
\end{figure*}

In this section we present the results from a number of simulations which are described in Table\,\ref{table:sims}. We investigate the influence of changing the diffusion coefficient, the density profile of the disk (which dictates the strength of the sink term), the energy of the CRs and the inclusion of an advection term. Before discussing the effect of varying the parameters, we present the results of a fiducial run. 

\subsection{Fiducial run}
\label{subsec:fid}
We perform a fiducial run (d030-adv0-p1-res01, shown in boldface in Table\,\ref{table:sims}) with the following parameters. The radial and vertical extent of the simulation is 10\,au. The CRs are injected at the magnetospheric radius, $r=0.07$\,au at a vertical height of 0.03\,au. The diffusion coefficient is 30$r_\mathrm{L}$, the energy of the CRs is 3\,GeV and advection is turned off. The list of parameters used are given in Table\,\ref{table:sims}. The resulting ionisation rate (using Eq.\,\ref{eq:ion-rate}) is plotted in Fig.\,\ref{fig:ionrate_d02} as a function of radius and height above the PPD.
\begin{table*}
\centering
\caption{List of parameters for simulations}
\begin{tabular}{@{}lllllllll@{}}
\hline
Run ID & $r$  & $z$  & E   & $D/c$ & $p$ & $dr(=dz)$ & Position of CRs & Advection term \\
       & [au] & [au] & [GeV] & [$r_\mathrm{L}$]  &     & [au]      & ($r,z$) [au] & On\,[km\,$\mathrm{s^{-1}}$]/Off \\
\hline
{\bf Varying $D$} \\
d380-adv0-p1-res01  & 10 & 10 & 3 & 380 & 1.0 & 0.01 & 0.07, 0.03 & Off\\
d300-adv0-p1-res01  & 10 & 10 & 3 & 300 & 1.0 & 0.01 & 0.07, 0.03 & Off\\
d120-adv0-p1-res01  & 10 & 10 & 3 & 120 & 1.0 & 0.01 & 0.07, 0.03 & Off\\
\textbf{d030-adv0-p1-res01} & 10 & 10 & 3 &30 & 1.0 & 0.01 & 0.07, 0.03 & Off\\
d003-adv0-p1-res01  & 10 & 10 & 3 & 3   & 1.0 & 0.01 &0.07, 0.03 & Off\\
\hline
{\bf Varying the density profile/sink term} \\
d030-adv0-p05-res01  & 10 & 10 & 3 & 30 & 0.5 & 0.01 & 0.07, 0.03 & Off\\
d030-adv0-p15-res01  & 10 & 10 & 3 & 30 & 1.5 & 0.01 & 0.07, 0.03 & Off\\
\hline
{\bf Advection} \\
d030-adv1-p1-res01   & 10 & 10 & 3 &30 & 1.0 & 0.01 & 0.07, 0.03 & 1.0\\
d030-adv10-p1-res01  & 10 & 10 & 3 &30 & 1.0 & 0.01 & 0.07, 0.03 & 10.0\\
d030-adv100-p1-res01 & 10 & 10 & 3 &30 & 1.0 & 0.01 & 0.07, 0.03 & 100.0\\
d030-adv10000-p1-res01 & 10 & 10 & 3 &30 & 1.0 & 0.01 & 0.07, 0.03 & $10^4$\\
\hline
{\bf Resolution study} \\
d030-adv0-p1-res005 & 5 & 5 & 3 &30 & 1.0 & 0.005 & 0.5, 0.5 & Off\\
d030-adv0-p1-res01  & 5 & 5 & 3 &30 & 1.0 & 0.01  & 0.5, 0.5 & Off\\
d030-adv0-p1-res02  & 5 & 5 & 3 &30 & 1.0 & 0.02  & 0.5, 0.5 & Off\\
d030-adv0-p1-res04  & 5 & 5 & 3 &30 & 1.0 & 0.04  & 0.5, 0.5 & Off\\
d030-adv0-p1-res08  & 5 & 5 & 3 &30 & 1.0 & 0.08  & 0.5, 0.5 & Off\\
\hline
{\bf Radial extent test}\\
d030-adv0-p1-res01-rad20 & 20 & 20 & 3 &30 & 1.0 & 0.01 & 0.07, 0.03 & Off\\ 
\hline 
\end{tabular}
\label{table:sims}
\end{table*}

The influence of the PPD in absorbing the CRs is visible in this plot. Fig.\,\ref{fig:ionfrac_metal}\,-\,\ref{fig:ionfrac_nometal} plot the ionisation fraction (from Eq.\,\ref{eq:ionfrac}) resulting from the ionisation rate in Fig.\,\ref{fig:ionrate_d02}. These plots represent the difference between using $\beta_\mathrm{r}$ (including metal ions) as the recombination rate in Eq.\,\ref{eq:ionfrac} and using $\beta_\mathrm{d}$ (excluding metal ions). 

\begin{figure}
\centering
 \includegraphics[width=0.5\textwidth]{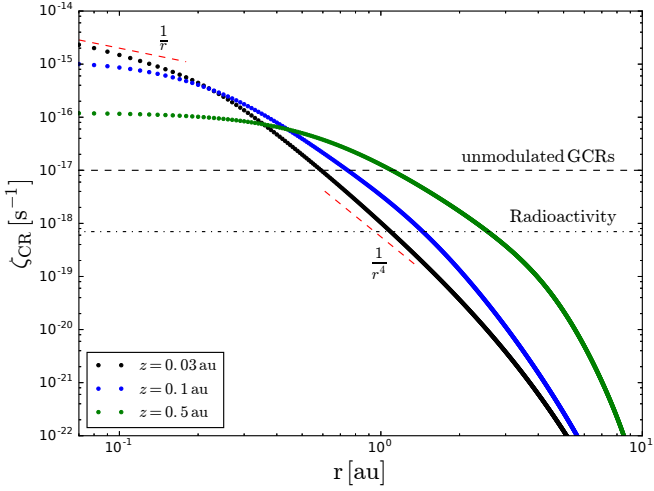}
  \caption{The ionisation rate, $\zeta_\mathrm{CR}$, for the fiducial run (d030-adv0-p1-res01) is plotted as a function of radius for different heights above the disk with the diffusion coefficient as $D/c= 30r_\mathrm{L}$.}
\label{fig:ionrate-d02}
\end{figure}
\begin{figure}
\centering
 \includegraphics[width=0.5\textwidth]{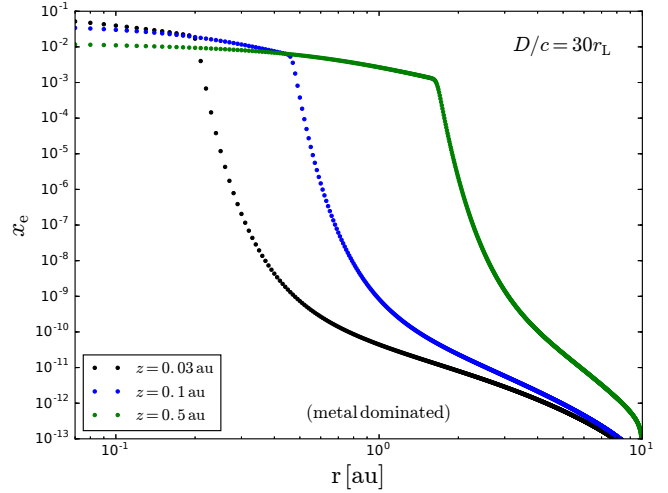}
  \caption{The ionisation fraction, $x_\mathrm{e}$, for the fiducial run (d030-adv0-p1-res01) is plotted as a function of radius for different heights above the disk with the diffusion coefficient as $D/c= 30r_\mathrm{L}$ and using $\beta_\mathrm{r}$. This corresponds to metal ions dominating as charge carriers in the plasma. The ionisation fraction when metal ions are absent is a factor of $3 \times 10^{-3}$ smaller. }
\label{fig:ionfrac-d02}
\end{figure}

To examine $\zeta_\mathrm{CR}$ more clearly near the midplane of the disk Fig.\,\ref{fig:ionrate-d02} plots $\zeta_\mathrm{CR}$ as a function of radius for $z=0.03$\,au (the dotted black line), the height at which the CRs are injected. The \emph{unmodulated} and \emph{unattenuated} ionisation rate from GCRs is denoted by the black dashed line in Fig.\,\ref{fig:ionrate-d02}. The ionisation rate from CRs originating from the young star is greater than an unattenuated GCR ionisation rate until $r\sim0.7\,$au at 0.03\,au above the midplane. The extremely low ionisation rate from radioactivity \citep{umebayashi_2009} is also plotted in Fig.\,\ref{fig:ionrate-d02} as the black dash-dotted line for comparison. For $r\gtrsim0.3\,$au, the ionisation rate is higher at greater heights above the disk as the attenuation of CRs is lower. Fig.\,\ref{fig:ionfrac-d02} plots the radial dependence of the ionisation fraction for d030-adv0-p1-res01 for different heights above the disk assuming that metals dominate.

Returning to Fig.\,\ref{fig:ionrate-d02}, in the innermost regions of the disk, at $z=0.03$\,au, the radial dependence of $\zeta_\mathrm{CR}$ is only very slightly steeper than the canonical $r^{-1}$ profile expected from diffusive processes in the absence of losses. Further out the radial dependence steepens to $\sim r^{-4}$, as indicated in the plot. The reason for this behaviour can be explained by the influence of the sink term. The mass density at the inner edge of the PPD is so high that the disk is absorbing the CRs faster than they can diffuse, thus curtailing the $r^{-1}$ dependence. Rather than dropping exponentially from the section of the disk described by an $r^{-1}$ dependence, the radial dependence is softer than an exponential drop-off because the density of the PPD decreases radially allowing the CRs that do diffuse through the densest regions of the disk to survive for longer. To confirm this interpretation and to test that the code does reproduce the expected $r^{-1}$ diffusive profile an artificial `low density' limit case was investigated which was described in Section\,\ref{appendix:low-density}. This finding also emphasises the potential importance of the density profile of the disk on the results which is discussed in Section\,\ref{subsec:rho}.

The blue and green dash-dotted lines in Fig.\,\ref{fig:ionrate-d02} correspond to $\zeta_\mathrm{CR}$ at greater vertical heights above the disk at $z=0.1\,$au and $z=0.5\,$au respectively. As the density of the PPD decreases sharply as a function of height above the disk, the CRs penetrate much further even at modest heights above the disk. Also, for greater heights above the disk the radial profile of the ionisation rate tends towards a $r^{-1}$ profile. 

Another parameter which controls the ability of the CRs to reach the outer regions of the disk is the diffusion coefficient which is described in Section\,\ref{subsec:d_vary}. As a result of the steep radial drop-off in $\zeta_\mathrm{CR}$, due to the high mass density of the PPD, increasing the luminosity of the CRs by an order of magnitude, for instance, would only result in the CRs penetrating out to $\sim 2$\,au at $z=0.03\,$au which is not a large effect. Thus, significantly increasing $L^\mathrm{CR}_*$ alone will not significantly increase $\zeta_\mathrm{CR}$ further out in the disk.

\subsection{Dependence on diffusion coefficient}
\label{subsec:d_vary}
\begin{figure}
\centering
 \includegraphics[width=0.5\textwidth]{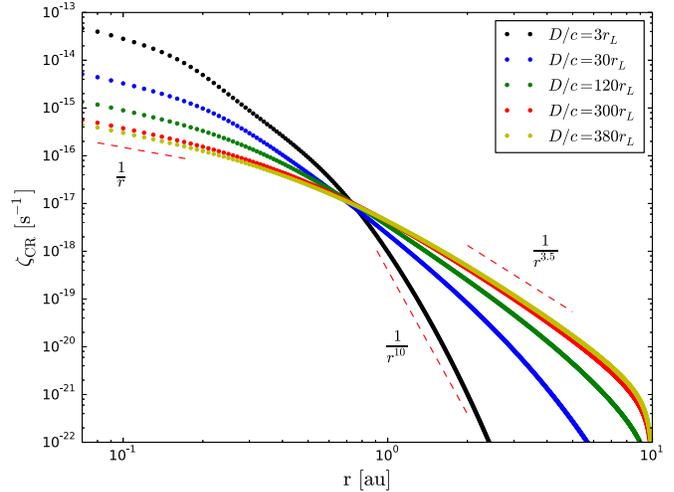}
  \caption{The ionisation rate, $\zeta_\mathrm{CR}$, is plotted as a function of radius for different diffusion coefficients.}
\label{fig:ionrate_comp}
\end{figure}

Next we consider the effect of varying the diffusion coefficient shown in Fig.\,\ref{fig:ionrate_comp}. The diffusion coefficient (given in units of $c$) is varied between $3-380r_\mathrm{L}$ with all other parameters held constant (details of the simulations are given in Table\,\ref{table:sims} as d003-adv0-p1-res01, d380-adv0-p1-res01, etc.). Varying the diffusion coefficient does not increase or decrease the distance over which stellar CRs dominate unattenuated GCRs because from Eq.\,\ref{eq:ion-rate} the ionisation rate is normalised to a particular density. Instead, as the diffusion coefficient increases (meaning that the CRs diffuse faster) $\zeta_\mathrm{CR}$ decreases in the inner regions of the disk, shown in Fig.\,\ref{fig:ionrate_comp}. The radial dependence moves more towards a $r^{-1}$ profile (see Fig.\,\ref{fig:ionrate_comp}) as the diffusion coefficient increases and thus the same ionisation rate is obtained at $\sim$0.7\,au. If the assumed CR luminosity was increased along with an increase in the diffusion coefficient then the ionisation rate would increase further out but varying the diffusion coefficient alone does not achieve this.

\subsection{Dependence on momentum/energy of CRs}
\label{subsec:results_energy}
Until now we have only considered 3\,GeV CRs but it is possible that lower and higher energy CRs are also present in the system. Higher energy CRs travel further than lower energy CRs before losing their energy since higher energy particles have larger diffusion coefficients. The diffusion coefficient varies as a function of energy according to Eq.\,\ref{eq:diff-coeff}. We include CRs with momenta between $0.1-300\,\mathrm{GeV/}c$ and weight the contribution from each momentum bin by the spectrum with the functional form described in Eq.\,\ref{eq:xi} and with $\alpha=2$. All other parameters are set to the fiducial values mentioned in Section\,\ref{subsec:fid}.

\begin{figure}
\centering
 \includegraphics[width=0.5\textwidth]{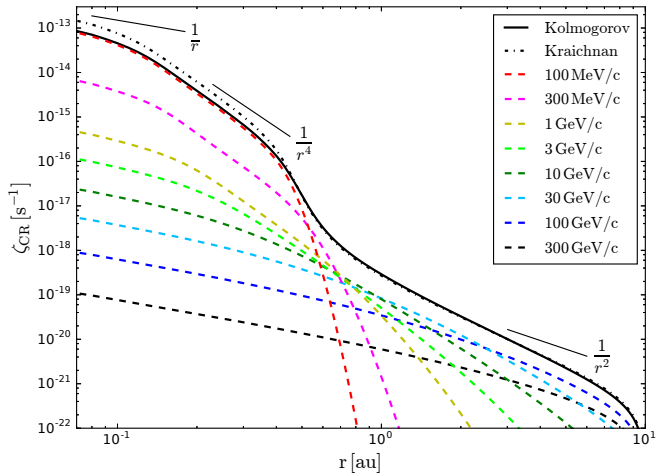}
  \caption{The ionisation rate, $\zeta_\mathrm{CR}$, resulting from a CR spectrum (including contributions from $0.1-100\,\mathrm{GeV}/c$ CRs) assuming an underlying Kolmogorov turbulence spectrum is plotted as a function of radius at $z=0.03\,$au, denoted by the solid black line. The result of instead assuming a Kraichnan turbulence spectrum is plotted as the dash-dotted black line. The individual contributions to $\zeta_\mathrm{CR}$, for the Kolmogorov case, from CRs of different momenta are given by the dashed lines. }
\label{fig:ionrate_energy}
\end{figure}

Fig.\,\ref{fig:ionrate_energy} plots $\zeta_\mathrm{CR}(r)$ obtained by including contributions from $0.1-300\,\mathrm{GeV}/c$ CRs at $z=0.03$\,au. The dotted black line corresponds to a Kolmogorov spectrum with $\gamma=5/3$. The dashed black line corresponds to a Kraichnan spectrum with $\gamma=3/2$. The net effect is negligible despite the different diffusion coefficients for a Kolmogorov or Kraichnan spectrum according to Eq.\,\ref{eq:dif-norm}. This is because the weighting factors from Eq.\,\ref{eq:xi} are more important in determining the overall spectrum.
The other coloured dashed lines correspond to the different energy components of the Kolmogorov spectrum (solid black line). Including an energy spectrum introduces more complexity to the radial dependence of the ionisation rate. In the inner region out to $\sim$1\,au, the low energy CR component dominates with a sharply decreasing slope. After 1\,au the spectrum hardens as the contribution from high energy CR dominates. Higher energy CRs can penetrate further in radius than low energy CRs which increases the ionisation rate at larger radii. Increasing the hardness (softness) of the CR energy spectrum would result in the contribution from high energy CRs becoming more (less) important in Fig.\,\ref{fig:ionrate_energy}. Changing the spectral index of the CR energy spectrum would reflect a change in the assumed underlying acceleration mechanism. Here, we chose $\alpha=2$ which is representative of a spectral index for a Fermi first order acceleration mechanism.

For this work we have assumed a constant loss rate per grammage, $\left( \frac{dE}{d\chi} \right)$, for different energy CRs. In reality, the energy deposition rate, $\left( \frac{dE}{d\chi} \right)$, depends upon the density of the medium and the energy of the CRs. However, for the sake of simplicity we have neglected these secondary effects here.

\subsection{Density profile of the PPD}
\label{subsec:rho}
\begin{figure}
\centering
 \includegraphics[width=0.5\textwidth]{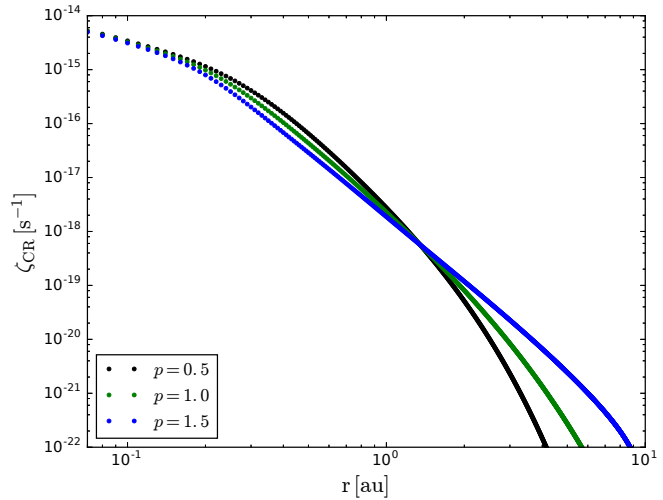}
  \caption{The ionisation rate as a function of radius for varying density profiles of the disk, ($p=0.5,1.0,1.5$) at $z=0.03\,$au.}
\label{fig:ionrate-slice-rho}
\end{figure}

We investigate the dependence of our results on the density profile of the PPD, given in Eq.\,\ref{eq:rho}. The radial dependence is determined by the parameter $p$. We vary this parameter between $0.5<p<1.5$ (Fig.\,\ref{fig:ionrate-slice-rho}) with all other values set to their fiducial values as described in Section\,\ref{subsec:fid}. The details of the simulations are given in Table\,\ref{table:sims}.
As $p$ increases and the density profile becomes steeper the CRs penetrate further out in the disk due to the decreased density in these regions. Although, at $\sim5$\,au there are orders of magnitude difference in $\zeta_\mathrm{CR}$ as a result of changing $p$, the ionisation rates for $r\lesssim 2\,$au are very similar. Altering $p$ does not decrease the density hugely in the innermost regions of the disk where the CRs are strongly attenuated which is why the ionisation rates differ so little.

The difference in the radial dependence of $\zeta_\mathrm{CR}$ at larger radii obtained by varying $p$ can be understood further by considering a simpler (unphysical) system. We consider the steady-state equation, Eq.\,\ref{eq:ncr-ss}, with a number of simplifications. We neglect the advective term in this analysis, assume spherical symmetry (meaning that the disk does not exist as such) and that the sink term is only a function of $r$. Hence, Eq.\,\ref{eq:ncr-ss} becomes
\begin{equation}
\frac{n_\mathrm{CR}}{\tau(r)} - \bm{\nabla}\cdot(D\bm{\nabla} n_\mathrm{CR}) = Q 
\end{equation}   
which can be solved analytically. If the sink term is artificially set to zero then $n_\mathrm{CR}(r) \propto r^{-1}$, whereas if $\tau(r)=C$, where $C$ is a constant, then the radial dependence of $n_\mathrm{CR}(r)$ becomes $n_\mathrm{CR}(r) \propto r^{-1}e^{-r/r_\tau}$, where $r_\tau$ dictates radially where the exponential cut-off begins. This case is similar to the test performed in Section\,\ref{appendix:low-density} where diffusion dominated over the sink term. If $\tau\propto r^a$ with $a\ge-2$ (which corresponds to varying $p$) then $n_\mathrm{CR}(r) \propto r^{-1}f(r)$ where the function $f(r)$ is not equal to an exponential cut-off and is instead much harder. This is a highly simplified system but it does illustrate how the sink term influences the radial dependence of $\zeta_\mathrm{CR}$. Note the disappearance of the exponential cut-off for the case of $p=1.5$ in Fig.\,\ref{fig:ionrate-slice-rho}. 

\subsection{Effect of the advection term}
\label{subsec:advection-results}
\begin{figure}
\centering
 \includegraphics[width=0.5\textwidth]{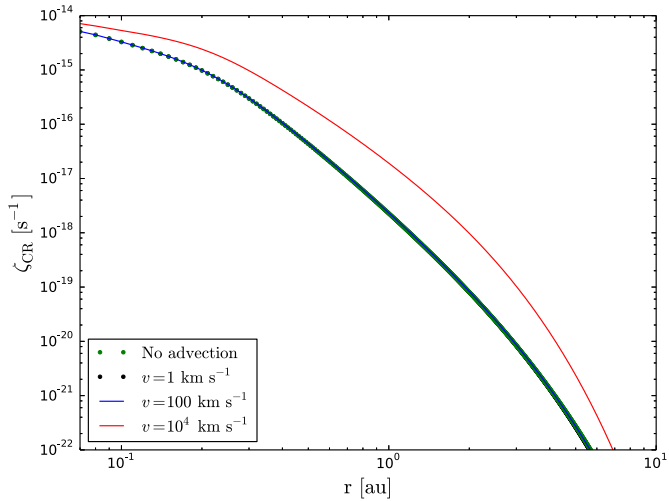}
  \caption{The ionisation rate as a function of radius including an advective wind, plotted at $z=0.03\,$au.}
\label{fig:advection}
\end{figure}

The last effect that we include is that of an advective wind, intended to mimic the effect that an MHD disk wind would have on the diffusion of CRs. As mentioned in Section\,\ref{subsubsec:adv} the advection speeds considered are in the range of $1-10^4\,\mathrm{km\,s^{-1}}$, the details of the simulations are given in Table\,\ref{table:sims}. The results of combining both diffusion and advection are shown in Fig.\,\ref{fig:advection}.

Fig.\,\ref{fig:advection} shows that for reasonable values of an advective wind between $1-100\mathrm{km\,s^{-1}}$ diffusion dominates and there is no significant difference in the ionisation rate at $z=0.03\,$au when advection is included. To check if faster wind speeds begin to have an effect we included an unphysically fast wind (for a low-mass young star) of $10^{4}\mathrm{km\,s^{-1}}$ which is the red line in Fig.\,\ref{fig:advection}. In this case the ionisation rate unexpectedly increases. This may occur because the wind  advects the CRs radially outwards away from the point of injection into a less dense area of the disk. Thus fewer CRs are lost initially in the dense inner region of the disk. This is perhaps an unphysical effect because in all of these simulations diffusion is still acting cospatially with the advective wind. It is unclear whether CRs entrained in an MHD wind would in fact be able to diffuse back towards the midplane of the disk in the way that is demonstrated in this simulation.

As mentioned above, only the most extreme (and unphysical) value for the advection speed ($v=10^4\,\mathrm{km\,s^{-1}}$) results in a significant change in the ionisation rate when advection is included. To confirm this finding we can compare the diffusive timescale, $t_\mathrm{diff}$, with the advective timescale, $t_\mathrm{adv}$ which are defined by
\begin{flalign}
&t_\mathrm{diff} = \frac{r^2}{D} \\
&t_\mathrm{adv} = \frac{r}{v}
\label{eq:timescale}
\end{flalign}

\begin{figure}
\centering
 \includegraphics[width=0.5\textwidth]{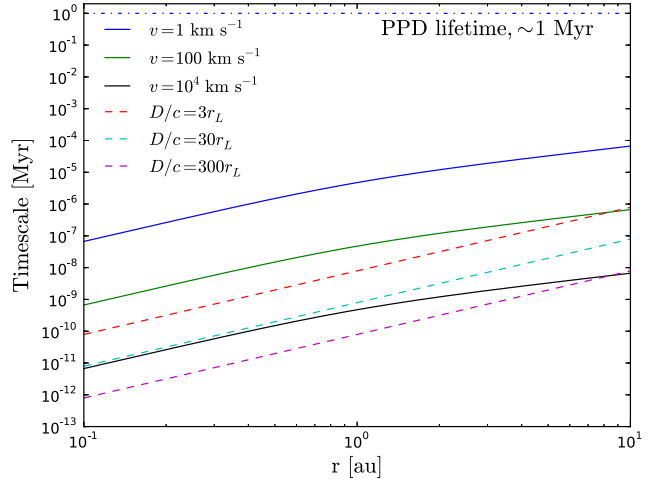}
  \caption{Illustration of the different timescales involved, the shortest of which will dictate the evolution of the system. These timescales are the diffusive (solid lines) and advective (dashed lines) timescales plotted for a range of values. The lifetime of the disk is also plotted (blue dash-dotted line) to ensure that it is much longer than the timescales of physical processes considered. Only the most extreme (unphysical) advective timescale corresponding to $v=10^4\mathrm{km\,s^{-1}}$ is competitive with any of the diffusive timescales within the computational domain but only for $r\lesssim1\,$au. For the case of $v=100\mathrm{km\,s^{-1}}$ advection also begins to dominate at $\sim10\,$au. The advective timescale does not increase linearly with radius because the radial component of $v$ decreases as a function of radius, as discussed in Section\,\ref{subsubsec:adv} and shown in Eq.\,\ref{eq:vr}.}
\label{fig:timescale}
\end{figure}

Fig.\,\ref{fig:timescale} plots a range of diffusive and advective timescales. A diffusion coefficient, divided by $c$, of $30r_\mathrm{L}$ and an unphysically fast advection speed for the system of $v=10^4\,\mathrm{km\,s^{-1}}$ is the only combination which allows advection to dominate at a radius considered in the simulations presented here. For the other parameters it is clear that advection does not dominate, comfirming the results plotted in Fig.\,\ref{fig:advection}. At large radii advection will eventually dominate over diffusion.
 
The approximate timescale for disk dispersal is denoted by the dashed blue line in Fig.\,\ref{fig:timescale}. Both the advective and diffusive timescales are much shorter than this timescale, indicating that the CRs would have reached steady state via diffusion or advection long before the disk dispersed.

\section{Discussion}
\label{sec:discussion}
\subsection{Comparison with observations}
As mentioned in Section\,\ref{sec:intro}, there are some observational hints suggesting that YSOs and the environments surrounding them are capable of accelerating particles up to at least $\sim$GeV. In this section we elaborate more on these observational results before comparing our results with ionisation rates expected from X-rays and discussing the specific case of TW Hya, a class II object for which an estimate of the ionisation rate is available \citep{cleeves-2015}.

\subsubsection{Observational hints of low energy cosmic rays}

\citet{ainsworth_2014} found evidence for non-thermal emission from the bow shock of a knot in the jet of DG Tau, a class II source. This non-thermal emission indicates the presence of $\sim$GeV cosmic rays (CRs) at the bow shock. Though located too far from the PPD to significantly ionise it, these observations do indicate that particle acceleration can occur in the jets of Class IIs.

 \citet{ceccarelli_2014} found an ionisation fraction higher than expected for OMC-2 FIR 4 (a Class 0 object) by examining the ratio of $\mathrm{HCO^+}$ and $\mathrm{N_2H^+}$. This suggests there is an, as of yet, unaccounted for source of ionisation in the system. 
 
Recently, \citet{padovani_2015,padovani_2016} investigated possible sites of acceleration for CRs in young protostellar systems (mainly relevant to Class 0 objects) via diffusive shock acceleration. They concluded that it was plausible to accelerate particles at the protostellar surface (for Class 0/I objects) and in jet shocks, whereas this paper focuses on Class II objects.
 
Finally, \citet{aresu-2014} examined the [O\,I] 63\,$\mu$m line from the PACS \citep[Photodetector Array Camera and Spectrometer,][]{poglitsch_2010} instrument on the \emph{Herschel} Space Observatory \citep{pilbratt_2010} for a sample of PPDs for which no jet emission had been detected. From \citet{aresu-2014}, this line was predicted by models \citep{woitke_2009,aresu_2012} to occur in the tenuous disk atmosphere at radii from approximately $10\,$au to $200\,$au. This region of the disk is thought to be directly exposed to stellar radiation and that the main sources of heating would be X-rays and FUV radiation. For the \emph{Herschel} sample of PPDs, \citet{aresu-2014} found no correlation between the [O\,I] 63\,$\mu$m line flux and X-ray luminosity of the CTTSs which would be expected if X-rays were responsible for the majority of the [O\,I] 63\,$\mu$m line flux emission. In this case, we postulate that another possible explanation for this is a hitherto unknown source of the heating being present in the system.

Detecting the ionisation rate via molecular tracers would be the best method of comparing with the results of the ionisation rate presented here. Observationally constraining the ionisation rate in PPDs is a particularly difficult task requiring high spatial resolution and sensitivity which has only become conceivable with the Atacama Large Millimeter Array (ALMA). To date, there are very few observations that have attempted to do this, however the (overall) ionisation rate for TW Hya is discussed below.

\subsubsection{Ionisation rate for TW Hya}
\citet{cleeves-2015} report an upper limit for the (overall) ionisation rate ($\zeta\lesssim10^{-19}\,\mathrm{s}^{-1}$) at the disk midplane, at all radii, for TW Hya by using observations of HCO$^+$ line emission from the Submillimeter Array and N$_2$H$^+$ from ALMA in conjunction with a grid of disk chemistry models. This is much lower than some of the results presented in this paper. There are a number of reasons that might explain this. The contribution from stellar CRs found in this paper is overall most significant at radii $\lesssim 10\,$au which ALMA and the SMA are not currently able to resolve. TW Hya may also be an unusual object: it is a $\sim3-10$\,Myr old T-Tauri star still hosting a PPD making it an unusually old disk \citep{barrado_2006,vacca_2011},  perhaps indicating that the presence of such an old disk is a result of low levels of ionisation which has impeded its evolution. From infrared excesses \citep{haisch-2001} and accretion rates \citep{fedele-2010} it can be inferred that most young stars do not possess disks at 10 Myr. TW Hya is often observed due to its close proximity, located at a distance of $55 \pm 9\,$pc \citep{webb-1999}, rather than because it is a prototypical class II object.  ALMA provides the possibility of constraining $\zeta_\mathrm{CR}$ for other PPDs which can be compared with our results.

\subsection{Comparison of X-ray with stellar CR ionisation}
\label{subsec:xray-cr}
This work focuses on the ionising effect of low energy cosmic rays from the young star itself. There is also a significant amount of X-ray ionisation from the young star, the effect of which has previously been studied \citep[][for example]{igea_1999}. Here, we compare these two sources of ionisation. To do this we use the parameterised fit of the results from \citet{igea_1999} given in \citet{bai_2009} to estimate the X-ray ionisation rate, $\zeta_\mathrm{XR}(r,z)$. We consider 5\,keV X-rays with a X-ray luminosity, $L_\mathrm{XR}$, of $1\times 10^{29}\mathrm{erg\,s^{-1}}$. The details of this parameterised fit from \citet{bai_2009} are given in Appendix\,\ref{appendix:xray}.

Near the midplane of the disk the CRs dominate over X-rays beyond $\sim 0.3\,$au where the ionisation rate is $\zeta\sim 10^{-16}\,\mathrm{s^{-1}}$, though the CRs themselves also drop-off significantly with increasing radius (shown in Fig.\,\ref{fig:xray-cr}). The X-rays propagate with a geometric dilution factor of $\sim r^{-2}$ until they become attenuated by the disk. We find that irrespective of the height above the disk that X-rays lead to higher ionisation rates closer to the star with the CRs dominating at larger radii. This occurs because the X-rays are lower in energy than the CRs and also because the CRs have a geometric dilution factor of $r^{-1}$ until they also become attenuated by the disk.

%% file: conclusions.tex
\section{Conclusions}
\label{sec:conclusions}

In this paper we investigated the ionising effect of low energy stellar CRs in PPDs by treating their propagation as a diffusive process. We found that for the chosen fiducial values the ionisation rate from 3\,GeV CRs is larger than that of unattenuated GCRs out to $\sim1\,$au at the disk midplane. Above the disk midplane the ionisation rate from stellar CRs continues to dominate over that of an unattenuated GCR ionisation rate at greater radii. If the luminosity of the CRs is larger than that considered here in combination with a higher diffusion coefficient then the resulting ionisation rate will accordingly increase.

\citet{cleeves-2013} showed that low energy GCRs are likely to be suppressed from the CTTS system due to the `T-Tauriosphere', the more powerful stellar analogue of the solar heliosphere. This removes an important source of ionisation from the disk in particular. Hence, this led us to consider the ionising influence of \emph{stellar} CRs injected at the inner edge of the PPD. The ionisation fraction in PPDs controls the level of coupling between the neutral species and the magnetic field via collisions with the charged species. This dictates the location and strength of non-ideal effects and the development of the MRI which may be responsible for angular momentum transport in PPDs. 

The canonical $r^{-1}$ dependence expected from diffusive processes is not recovered in the simulations presented here for the majority of the diffusion coefficients investigated because of the large densities in the inner regions of PPDs which attenuates the CRs. As the diffusion coefficient of the CRs is increased the radial dependence converges towards a $r^{-1}$ dependence.

Introducing an energy spectrum of CRs leads to multiple power law components being present in the radial dependence of the ionisation rate. In the inner regions of the disk the contribution from low energy CRs dominates with a steep power law dependence and the contribution from higher energy CRs becomes more important further out than 1\,au with a harder spectrum.

The influence of an MHD wind in deflecting the CRs was investigated by including an advection term in the transport equation. The advection timescales for a range of advection speeds ($1-100\,\mathrm{km\,s^{-1}}$) are much longer than the diffusive timescales considered meaning that the advection term has little effect. Advection would become relevant if the diffusion coefficient is much smaller than what was considered to be physically reasonable in this paper. Advection could also be competitive for winds with $v\gg100\,\mathrm{km\,s^{-1}}$, as demonstrated by the case of $v=10^4\,\mathrm{km\,s^{-1}}$, but as discussed in Section\,\ref{subsec:advection-results} this is unphysically fast for disk winds from PPDs. 

The ability to treat the propagation of CRs as diffusive relies on the presence of turbulent magnetic fields. We have assumed that the inner regions of the PPD are turbulent due to thermal ionisation and the onset of the MRI. The additional ionisation source of stellar CRs would allow regions further out in the disk to become MRI active. These regions containing turbulent magnetic fields would then act to trap more CRs, further enhancing the ionisation rates. The level of turbulence in the disk affects the diffusion coefficients. Higher levels of turbulence lead to lower diffusion coefficients. In this paper we have neglected the feedback of the disk's potentially time variable magnetic field configuration on the propagation of the stellar CRs. The X-ray luminosity of YSOs also shows variability which would imply that the production of CRs is perhaps similarly variable which is not taken into account here.

The energy loss rate used in Section\,\ref{subsubsec:sink-term} represents an average of different energy loss processes. Notably, if it is decomposed there is a contribution from neutrons resulting from $pp$ interactions, as discussed in \citet{kataoka_2016}. It would perhaps be of interest to investigate the ionising effect of the neutrons separately. 1 (100) GeV neutrons have a lifetime of $\sim10^3\,(10^5)\,$s allowing them to travel $\sim2\,(200)$au before decaying, irrespective of the presence of a magnetic field. This also implies that PPDs would be dim emitters of $\sim$GeV neutrinos, again as a result of $pp$ interactions. In this paper we have assumed that the CRs are produced at the magnetospheric radius. Stochastic acceleration however could also be a relevant process throughout turbulent parts of the disk and would introduce a delocalised source term into the problem.

Lastly, we note that the presence of low energy stellar CRs would have a significant influence on the ionisation rate in the inner parts of PPDs and could alleviate to some extent the issue of very low ionisation rates in PPDs due to GCRs being potentially excluded from the system by the `T-Tauriosphere'.

%% file: acknowledgements.tex
\section*{Acknowledgements}
D.R.L. and T.P.R. acknowledge support from Science Foundation Ireland under grant 11/RFP/AST3331. D.R.L. would like to thank Antonella Natta and Malcolm Walmsley for many helpful discussions and for encouraging her in this direction. We would like to thank the anonymous referee for helpful comments which improved the manuscript.

%% file: appendix.tex
\section{CR transport equation as a difference equation}
\label{appendix:dif-eq}
\begin{details}
Before giving Eq.\,\ref{eq:ncr} as a difference equation we can first express it in cylindrical coordinates as
\begin{eqnarray}
\frac{\partial n_\mathrm{CR}}{\partial t} = \frac{1}{r}\frac{\partial}{\partial r}\bigg(rD\frac{\partial n_\mathrm{CR}}{\partial r}\bigg) + \frac{\partial}{\partial z}\bigg(D\frac{\partial n_\mathrm{CR}}{\partial z}\bigg) \nonumber \\
- \frac{1}{\tau}n_\mathrm{CR}  - v_r\frac{1}{r}\frac{\partial}{\partial r}(rn_\mathrm{CR}) - v_z\frac{\partial n_\mathrm{CR}}{\partial z} + Q
\label{eq:ncr1}
\end{eqnarray}
Writing Eq.\,\ref{eq:ncr1} in terms of the Lax-Wendroff scheme gives
\begin{eqnarray}
\frac{\partial n_\mathrm{CR}}{\partial t} = \frac{1}{r}\frac{\partial}{\partial r}\bigg(rD\frac{\partial n_\mathrm{CR}}{\partial r}\bigg) + \frac{\partial}{\partial z}\bigg(D\frac{\partial n_\mathrm{CR}}{\partial z}\bigg) \nonumber \\
- \frac{1}{\tau}n_\mathrm{CR}  - v_r\frac{1}{r}\frac{\partial}{\partial r}(rn_\mathrm{CR}) - v_z\frac{\partial n_\mathrm{CR}}{\partial z} + Q
\label{eq:ncr1}
\end{eqnarray}
\end{details}

\noindent Eq.\,\ref{eq:ncr} can be written out as a difference equation using the Lax-Wendroff scheme for the advective term on a discrete cylindrical grid of points $(r_i,z_j,t_n)$ with
\begin{equation}
r_i = i\Delta r, \hspace{2mm} z_j = i\Delta z , \hspace{2mm} t_n = n\Delta t
\end{equation}
\noindent where $\Delta r$ ($\Delta z$) is the radial (vertical) grid spacing, in this case $\Delta r= \Delta z$ and $\Delta t$ is the timestep. The numerical approximation of $n_\mathrm{CR}$ at position and time $(r_i,z_j,t_n)$ is given as $u^n_{i,j}$. From now on, upper indices refer to time and the lower indices refer to space. We would like to express the next timestep, $u^{n+1}_{i,j}$, in terms of the previous timestep, $u^n_{i,j}$ which is given as follows
\begin{eqnarray}
&u^{n+1}_{i,j}& = u^{n}_{i,j} \nonumber \\
&+& \frac{\Delta t}{4r_i\Delta r^2}\Big[(r_{i+1}+r_{i})(D_{r,i+1,j} +D_{r,i,j})(u^n_{i+1,j}-u^n_{i,j}) \nonumber \\
&-& (r_{i}+r_{i-1})(D_{r,i,j} +D_{r,i-1,j})(u^n_{i,j}-u^n_{i-1,j})\Big]  \nonumber \\
&+&\frac{\Delta t}{2\Delta z^2}\Big[(D_{z,i,j+1} +D_{z,i,j})(u^n_{i,j+1} - u^n_{i,j}) \nonumber \\
&-& (D_{z,i,j} +D_{z,i,j-1})(u^n_{i,j} - u^n_{i,j-1})\Big]  \nonumber \\
&-&  \Delta tc_{i,j}u^n_{i,j} \nonumber \\
&-& \Delta t v^n_{r,i,j}\frac{u^n_{i+1,j}-u^n_{i-1,j}}{2\Delta r} \nonumber \\
&+& \frac{\Delta t^2}{2}\Big[ -\frac{v_{r,i,j}^n v_{r,i,j}^n}{2r\Delta r}(u^n_{i+1,j} - u^n_{i-1,j}) \nonumber \\
&+& \frac{v^n_{r,i,j}v^n_{r,i,j}(u^n_{i+1,j} - 2u^n_{i,j}+  u^n_{i-1,j})}{\Delta r^2}\Big] \nonumber \\
 &-&\Delta t v^n_{z,i,j}\frac{u^n_{i,j+1}-u^n_{i,j-1}}{2\Delta z} \nonumber \\
 &+& \frac{\Delta t^2}{2}\Big[\frac{v_{z,i,j}^nv_{z,i,j}^n}{\Delta z^2}\left(u^n_{i,j+1} - 2u^n_{i,j} + u^n_{i,j-1} \right)  \Big] \nonumber \\
\end{eqnarray}
In the above equation $D_{r,i,j} (v^n_{r,i,j})$ and $D_{z,i,j} (v^n_{z,i,j})$ are the components of the diffusion coefficient (velocity) in the radial and vertical direction, respectively. We have assumed in this paper that $D_{r,i,j}=D_{z,i,j}$. The loss rate, $1/\tau(r_i,z_j)$ is represented by $c_{i,j}$ and
\begin{equation}
\frac{\partial u}{\partial r} \biggr\rvert_{i+1/2,j} = \frac{u_{i+1,j} - u_{i,j}}{\Delta r};\,\,\, u^n_{i+1/2,j} = \frac{u^n_{i+1,j} +u^n_{i,j}}{2} 
\end{equation}
and $\dfrac{\partial u}{\partial z} \biggr\rvert_{i+1/2,j}$ and $u^n_{i,j+1/2}$ are defined in a similar way.

\section{Numerical convergence test}
\label{appendix:res}
\begin{figure}
\centering
 \includegraphics[width=0.5\textwidth]{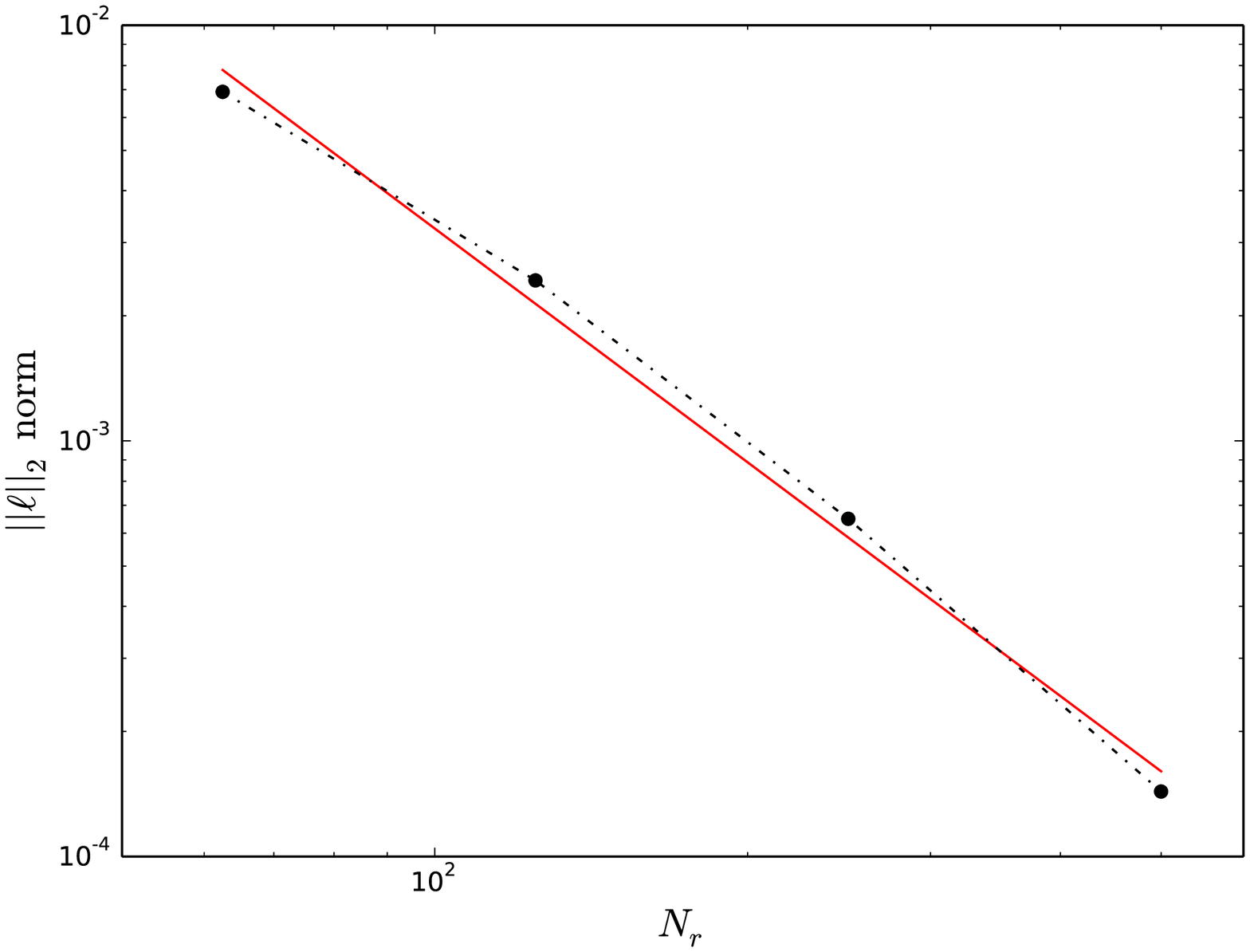}
  \caption[$||\ell||_2$ norm plotted as a function of resolution]{$||\ell||_2$ norm plotted as a function of resolution where $N_r$ is the number of grid zones in the radial direction. In this case, the number of cells in the vertical direction is also equal to $N_r$.}
\label{fig:l2-norm}
\end{figure}

To test that the code has the expected rate of convergence we examine the $||\ell||_2$ norm as a function of resolution, shown in Fig.\,\ref{fig:l2-norm}. The $||\ell||_2$ norm is defined as
\begin{equation}
||\ell||_2 = \frac{1}{n}\sqrt{\sum^n_{i=0}|x_{i,1}-x_{i,2}|^2}
\end{equation}
where the index $i$ indicates the spatial position and the indices $1,2$ correspond to two different resolution simulations. When plotted on a log-log scale the least-squared fitted slope is -1.86, in comparison to -2 which is the expected result for a second order accurate scheme. This verifies that the code is converging numerically. 

To perform this test over a range of resolutions the radius and height of the disk are both limited to 5\,au. The simulations are all stopped at the same time so that the results are comparable, rather than requiring them to reach a steady-state. The injection site of the CRs is moved to $r,z=(0.5\mathrm{au},0.5\mathrm{au})$ so that for all simulations the CRs would not be injected near a boundary cell. The parameters for these simulations are also given in Table\,\ref{table:sims}.

\section{X-ray ionisation}
\label{appendix:xray}
This section details the method used to estimate the ionisation rate from X-rays which is compared with CR ionisation in Section\,\ref{subsec:xray-cr}. The ionisation rate from X-rays, $\zeta_\mathrm{XR}$, produced in the vicinity of a CTTS can be obtained from Monte Carlo radiative transfer calculations such as in \citet{igea_1999}. These calculations have been fitted by \citet{bai_2009}, given by their Eq.\,21 and adapted here as,
\begin{eqnarray}
\zeta_\mathrm{XR} = \left(\frac{L_\mathrm{XR}}{L_{29}}\right)\left( \frac{r_\mathrm{in}}{R}\right)^{2.2} \bigg[\zeta_1\Big[e^{-(N_\mathrm{H_1}/N_1)^\alpha} + e^{-(N_\mathrm{H_2}/N_1)^\alpha}\Big] \nonumber \\
+ \zeta_2\Big[e^{-(N_\mathrm{H_1}/N_2)^\beta} + e^{-(N_\mathrm{H_2}/N_2)^\beta}\Big]\bigg] 
\end{eqnarray}

\noindent where $L_\mathrm{XR}$ is the X-ray luminosity of the star normalised by $L_{29}=1\times 10^{29}\mathrm{erg\,s^{-1}}$, $R$ is the cylindrical radius to the star and $r_\mathrm{in}$ is the inner edge of the PPD, similar to that described in Eq.\,\ref{eq:rho}. $N_\mathrm{H_1}$ and $N_\mathrm{H_2}$ are the vertical column densities above and below the disk, respectively, for a particular location in the disk, where $N_1$ and $N_2$ act as normalising column densities.

Here we have assumed that $L_\mathrm{XR}=1\times 10^{29}\mathrm{erg\,s^{-1}}$. We have normalised the X-ray luminosity to the inner edge of the disk, $r_\mathrm{in}$. We have chosen the X-ray energy to be 5\,keV but, as noted in \citet{bai_2009}, there is only a weak dependence of the results from \citet{igea_1999} on X-ray energy. For this energy, $\zeta_1 = 4.0\times 10^{-12}\,\mathrm{s^{-1}}$, $N_1 = 3.0\times 10^{21}\mathrm{cm^{-2}}$ and $\alpha=0.5$. The two exponentials multiplied by $\zeta_1$ correspond to contributions to $\zeta_\mathrm{XR}$ from the absorption of X-rays coming from above and below the disk. The two exponentials multiplied by $\zeta_2$ similarly represent a contribution to $\zeta_\mathrm{XR}$ from the scattering of X-rays. For these terms, $\zeta_2=2.0\times 10^{-15} \mathrm{s^{-1}}$, $N_2 = 1.0\times 10^{24}\mathrm{cm^{-2}}$ and $\beta=0.7$.  

\begin{figure}
\centering
 \includegraphics[width=0.5\textwidth]{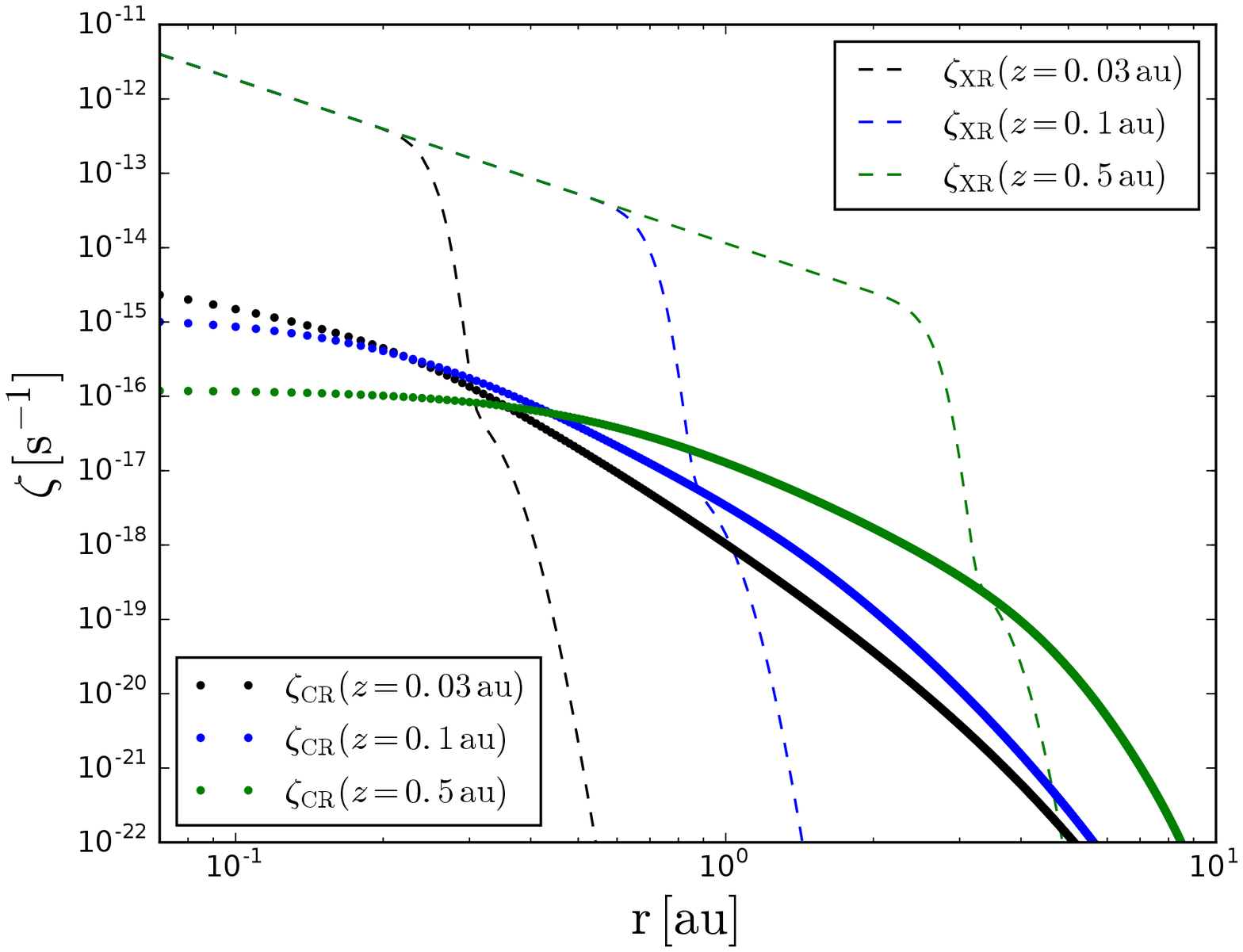}
  \caption{This plot compares the estimated ionisation rate as a function of radius from low energy stellar cosmic rays ($\zeta_\mathrm{CR}$) and X-rays ($\zeta_\mathrm{XR}$) for different heights above the disk.}
\label{fig:xray-cr}
\end{figure}

Fig.\,\ref{fig:xray-cr} shows the ionisation rate from low energy CRs and X-rays for a number of heights above the disk midplane. The dashed lines correspond to $\zeta_\mathrm{XR}$, at various heights above the disk, which decrease as $R^{-2.2}$ until they reach the disk and become attenuated quickly. For $z=0.03\,$au the CRs dominate over X-rays at radii greater than $\sim 0.3$\,au. The X-rays begin to dominate further out for greater heights above the disk illustrating that the CRs are most competitive as a source of ionisation in the denser regions of the PPD.

%% file: cr_clean.bbl
\begin{thebibliography}{}

\bibitem[\protect\citeauthoryear{{Ackermann}, {Ajello}, {Albert}, {Allafort},
  {Baldini}, {Barbiellini}, {Bastieri}, {Fermi LAT Collaboration} \& {et.
  al}}{{Ackermann} et~al.}{2014}]{ackermann_2014}
{Ackermann} M.,  {Ajello} M.,  {Albert} A.,  {Allafort} A.,  {Baldini} L.,
  {Barbiellini} G.,  {Bastieri} D.,  {Fermi LAT Collaboration}   {et. al} 2014,
  \apj, 787, 15

\bibitem[\protect\citeauthoryear{{Adams}, {Lada} \& {Shu}}{{Adams}
  et~al.}{1987}]{adams_1987}
{Adams} F.~C.,  {Lada} C.~J.,    {Shu} F.~H.,  1987, \apj, 312, 788

\bibitem[\protect\citeauthoryear{{Aguilar}, {Aisa}, {Alpat}, {Alvino},
  {Ambrosi}, {Andeen}, {Arruda}, {Attig}, {Azzarello}, {Bachlechner} \& et
  al.}{{Aguilar} et~al.}{2015}]{aguilar_2015}
{Aguilar} M.,  {Aisa} D.,  {Alpat} B.,  {Alvino} A.,  {Ambrosi} G.,  {Andeen}
  K.,  {Arruda} L.,  {Attig} N.,  {Azzarello} P.,  {Bachlechner} A.,    et al.
  2015, Physical Review Letters, 114, 171103

\bibitem[\protect\citeauthoryear{{Ainsworth}, {Scaife}, {Ray}, {Taylor},
  {Green} \& {Buckle}}{{Ainsworth} et~al.}{2014}]{ainsworth_2014}
{Ainsworth} R.~E.,  {Scaife} A.~M.~M.,  {Ray} T.~P.,  {Taylor} A.~M.,  {Green}
  D.~A.,    {Buckle} J.~V.,  2014, \apjl, 792, L18

\bibitem[\protect\citeauthoryear{{Ajello}, {Albert}, {Allafort}, {Baldini},
  {Barbiellini}, {Bastieri}, {Bellazzini}, {Bissaldi}, {Bonamente}, {Brandt},
  {Bregeon}, {Brigida}, {Bruel} \& {et. al}}{{Ajello}
  et~al.}{2014}]{ajello_2014}
{Ajello} M.,  {Albert} A.,  {Allafort} A.,  {Baldini} L.,  {Barbiellini} G.,
  {Bastieri} D.,  {Bellazzini} R.,  {Bissaldi} E.,  {Bonamente} E.,  {Brandt}
  T.~J.,  {Bregeon} J.,  {Brigida} M.,  {Bruel} P.,    {et. al} 2014, \apj,
  789, 20

\bibitem[\protect\citeauthoryear{{Andr{\'e}}, {Ward-Thompson} \&
  {Barsony}}{{Andr{\'e}} et~al.}{1993}]{andre_1993}
{Andr{\'e}} P.,  {Ward-Thompson} D.,    {Barsony} M.,  1993, \apj, 406, 122

\bibitem[\protect\citeauthoryear{{Andrews} \& {Williams}}{{Andrews} \&
  {Williams}}{2005}]{andrews_2005}
{Andrews} S.~M.,  {Williams} J.~P.,  2005, \apj, 631, 1134

\bibitem[\protect\citeauthoryear{{Aresu}, {Kamp}, {Meijerink}, {Spaans},
  {Vicente}, {Podio}, {Woitke}, {Menard}, {Thi}, {G{\"u}del} \&
  {Liebhart}}{{Aresu} et~al.}{2014}]{aresu-2014}
{Aresu} G.,  {Kamp} I.,  {Meijerink} R.,  {Spaans} M.,  {Vicente} S.,  {Podio}
  L.,  {Woitke} P.,  {Menard} F.,  {Thi} W.-F.,  {G{\"u}del} M.,    {Liebhart}
  A.,  2014, \aap, 566, A14

\bibitem[\protect\citeauthoryear{{Aresu}, {Meijerink}, {Kamp}, {Spaans}, {Thi}
  \& {Woitke}}{{Aresu} et~al.}{2012}]{aresu_2012}
{Aresu} G.,  {Meijerink} R.,  {Kamp} I.,  {Spaans} M.,  {Thi} W.-F.,
  {Woitke} P.,  2012, \aap, 547, A69

\bibitem[\protect\citeauthoryear{{Bai} \& {Goodman}}{{Bai} \&
  {Goodman}}{2009}]{bai_2009}
{Bai} X.-N.,  {Goodman} J.,  2009, \apj, 701, 737

\bibitem[\protect\citeauthoryear{{Balbus} \& {Hawley}}{{Balbus} \&
  {Hawley}}{1991}]{balbus_1991}
{Balbus} S.~A.,  {Hawley} J.~F.,  1991, \apj, 376, 214

\bibitem[\protect\citeauthoryear{{Barrado Y Navascu{\'e}s}}{{Barrado Y
  Navascu{\'e}s}}{2006}]{barrado_2006}
{Barrado Y Navascu{\'e}s} D.,  2006, \aap, 459, 511

\bibitem[\protect\citeauthoryear{{Blandford} \& {Payne}}{{Blandford} \&
  {Payne}}{1982}]{blandford_1982}
{Blandford} R.~D.,  {Payne} D.~G.,  1982, \mnras, 199, 883

\bibitem[\protect\citeauthoryear{{Ceccarelli}, {Dominik}, {L{\'o}pez-Sepulcre},
  {Kama}, {Padovani}, {Caux} \& {Caselli}}{{Ceccarelli}
  et~al.}{2014}]{ceccarelli_2014}
{Ceccarelli} C.,  {Dominik} C.,  {L{\'o}pez-Sepulcre} A.,  {Kama} M.,
  {Padovani} M.,  {Caux} E.,    {Caselli} P.,  2014, \apjl, 790, L1

\bibitem[\protect\citeauthoryear{{Chiang} \& {Goldreich}}{{Chiang} \&
  {Goldreich}}{1997}]{chiang_1997}
{Chiang} E.~I.,  {Goldreich} P.,  1997, \apj, 490, 368

\bibitem[\protect\citeauthoryear{{Cleeves}, {Adams} \& {Bergin}}{{Cleeves}
  et~al.}{2013}]{cleeves-2013}
{Cleeves} L.~I.,  {Adams} F.~C.,    {Bergin} E.~A.,  2013, \apj, 772, 5

\bibitem[\protect\citeauthoryear{{Cleeves}, {Bergin}, {Qi}, {Adams} \&
  {{\"O}berg}}{{Cleeves} et~al.}{2015}]{cleeves-2015}
{Cleeves} L.~I.,  {Bergin} E.~A.,  {Qi} C.,  {Adams} F.~C.,    {{\"O}berg}
  K.~I.,  2015, \apj, 799, 204

\bibitem[\protect\citeauthoryear{{Crutcher}}{{Crutcher}}{2012}]{crutcher_2012}
{Crutcher} R.~M.,  2012, \araa, 50, 29

\bibitem[\protect\citeauthoryear{{Decker}, {Krimigis}, {Roelof} \&
  {Hill}}{{Decker} et~al.}{2008}]{decker_2008}
{Decker} R.~B.,  {Krimigis} S.~M.,  {Roelof} E.~C.,    {Hill} M.~E.,  2008, in
  {Li} G.,  {Hu} Q.,  {Verkhoglyadova} O.,  {Zank} G.~P.,  {Lin} R.~P.,
  {Luhmann} J.,  eds, American Institute of Physics Conference Series Vol.~1039
  of American Institute of Physics Conference Series, {Particle Acceleration at
  the Termination Shock: Voyager 1 and 2 Observations}.
pp 349--354

\bibitem[\protect\citeauthoryear{{Drury}}{{Drury}}{2012}]{drury-2012}
{Drury} L.~O.~.,  2012, Astroparticle Physics, 39, 52

\bibitem[\protect\citeauthoryear{{Fedele}, {van den Ancker}, {Henning},
  {Jayawardhana} \& {Oliveira}}{{Fedele} et~al.}{2010}]{fedele-2010}
{Fedele} D.,  {van den Ancker} M.~E.,  {Henning} T.,  {Jayawardhana} R.,
  {Oliveira} J.~M.,  2010, \aap, 510, A72

\bibitem[\protect\citeauthoryear{{Feigelson} \& {Montmerle}}{{Feigelson} \&
  {Montmerle}}{1999}]{feigelson_1999}
{Feigelson} E.~D.,  {Montmerle} T.,  1999, \araa, 37, 363

\bibitem[\protect\citeauthoryear{{Frank}, {Ray}, {Cabrit}, {Hartigan}, {Arce},
  {Bacciotti}, {Bally}, {Benisty}, {Eisl{\"o}ffel}, {G{\"u}del}, {Lebedev},
  {Nisini} \& {Raga}}{{Frank} et~al.}{2014}]{prpl-2014}
{Frank} A.,  {Ray} T.~P.,  {Cabrit} S.,  {Hartigan} P.,  {Arce} H.~G.,
  {Bacciotti} F.,  {Bally} J.,  {Benisty} M.,  {Eisl{\"o}ffel} J.,  {G{\"u}del}
  M.,  {Lebedev} S.,  {Nisini} B.,    {Raga} A.,  2014, Protostars and Planets
  VI, pp 451--474

\bibitem[\protect\citeauthoryear{{Fromang}, {Terquem} \& {Balbus}}{{Fromang}
  et~al.}{2002}]{fromang_2002}
{Fromang} S.,  {Terquem} C.,    {Balbus} S.~A.,  2002, \mnras, 329, 18

\bibitem[\protect\citeauthoryear{{Gammie}}{{Gammie}}{1996}]{gammie_1996}
{Gammie} C.~F.,  1996, \apj, 457, 355

\bibitem[\protect\citeauthoryear{{Gressel}, {Turner}, {Nelson} \&
  {McNally}}{{Gressel} et~al.}{2015}]{gressel_2015}
{Gressel} O.,  {Turner} N.~J.,  {Nelson} R.~P.,    {McNally} C.~P.,  2015,
  \apj, 801, 84

\bibitem[\protect\citeauthoryear{{Haisch} Jr., {Lada} \& {Lada}}{{Haisch}
  et~al.}{2001}]{haisch-2001}
{Haisch} Jr. K.~E.,  {Lada} E.~A.,    {Lada} C.~J.,  2001, \apjl, 553, L153

\bibitem[\protect\citeauthoryear{{Hartmann}}{{Hartmann}}{2009}]{hartmann_2009}
{Hartmann} L.,  2009, {Accretion Processes in Star Formation: Second Edition}.
Cambridge University Press

\bibitem[\protect\citeauthoryear{{Hawley} \& {Balbus}}{{Hawley} \&
  {Balbus}}{1991}]{hawley_1991}
{Hawley} J.~F.,  {Balbus} S.~A.,  1991, \apj, 376, 223

\bibitem[\protect\citeauthoryear{{Igea} \& {Glassgold}}{{Igea} \&
  {Glassgold}}{1999}]{igea_1999}
{Igea} J.,  {Glassgold} A.~E.,  1999, \apj, 518, 848

\bibitem[\protect\citeauthoryear{{Johnstone}, {Jardine}, {Gregory}, {Donati} \&
  {Hussain}}{{Johnstone} et~al.}{2014}]{johnstone_2014}
{Johnstone} C.~P.,  {Jardine} M.,  {Gregory} S.~G.,  {Donati} J.-F.,
  {Hussain} G.,  2014, \mnras, 437, 3202

\bibitem[\protect\citeauthoryear{{Jokipii}}{{Jokipii}}{1966}]{jokipii-1966}
{Jokipii} J.~R.,  1966, \apj, 146, 480

\bibitem[\protect\citeauthoryear{{Kataoka} \& {Sato}}{{Kataoka} \&
  {Sato}}{2016}]{kataoka_2016}
{Kataoka} R.,  {Sato} T.,  2016, ArXiv e-prints

\bibitem[\protect\citeauthoryear{{Kraichnan}}{{Kraichnan}}{1965}]{kraichnan_1965}
{Kraichnan} R.~H.,  1965, \pof, 8, 1385

\bibitem[\protect\citeauthoryear{{Lada}}{{Lada}}{1987}]{lada_1987}
{Lada} C.~J.,  1987, in {Peimbert} M.,  {Jugaku} J.,  eds, Star Forming Regions
  Vol.~115 of IAU Symposium, {Star formation - From OB associations to
  protostars}.
pp 1--17

\bibitem[\protect\citeauthoryear{Lax \& Wendroff}{Lax \&
  Wendroff}{1960}]{lax_1960}
Lax P.,  Wendroff B.,  1960, Communications on Pure and Applied Mathematics,
  13, 217

\bibitem[\protect\citeauthoryear{{Levy}}{{Levy}}{1978}]{levy_1978}
{Levy} E.~H.,  1978, \nat, 276, 481

\bibitem[\protect\citeauthoryear{{Longair}}{{Longair}}{2011}]{longair_2011}
{Longair} M.~S.,  2011, {High Energy Astrophysics}

\bibitem[\protect\citeauthoryear{{Manara}, {Testi}, {Natta}, {Rosotti},
  {Benisty}, {Ercolano} \& {Ricci}}{{Manara} et~al.}{2014}]{manara_2014}
{Manara} C.~F.,  {Testi} L.,  {Natta} A.,  {Rosotti} G.,  {Benisty} M.,
  {Ercolano} B.,    {Ricci} L.,  2014, \aap, 568, A18

\bibitem[\protect\citeauthoryear{{Mewaldt}, {Cohen}, {Labrador}, {Leske},
  {Mason}, {Desai}, {Looper}, {Mazur}, {Selesnick} \& {Haggerty}}{{Mewaldt}
  et~al.}{2005}]{mewaldt_2005}
{Mewaldt} R.~A.,  {Cohen} C.~M.~S.,  {Labrador} A.~W.,  {Leske} R.~A.,  {Mason}
  G.~M.,  {Desai} M.~I.,  {Looper} M.~D.,  {Mazur} J.~E.,  {Selesnick} R.~S.,
   {Haggerty} D.~K.,  2005, Journal of Geophysical Research (Space Physics),
  110, A09S18

\bibitem[\protect\citeauthoryear{{Padovani}, {Hennebelle}, {Marcowith} \&
  {Ferri{\`e}re}}{{Padovani} et~al.}{2015}]{padovani_2015}
{Padovani} M.,  {Hennebelle} P.,  {Marcowith} A.,    {Ferri{\`e}re} K.,  2015,
  \aap, 582, L13

\bibitem[\protect\citeauthoryear{{Padovani}, {Marcowith}, {Hennebelle} \&
  {Ferri{\`e}re}}{{Padovani} et~al.}{2016}]{padovani_2016}
{Padovani} M.,  {Marcowith} A.,  {Hennebelle} P.,    {Ferri{\`e}re} K.,  2016,
  \aap, 590, A8

\bibitem[\protect\citeauthoryear{{Peres}, {Orlando}, {Reale}, {Rosner} \&
  {Hudson}}{{Peres} et~al.}{2000}]{peres-2000}
{Peres} G.,  {Orlando} S.,  {Reale} F.,  {Rosner} R.,    {Hudson} H.,  2000,
  \apj, 528, 537

\bibitem[\protect\citeauthoryear{{Perez-Becker} \& {Chiang}}{{Perez-Becker} \&
  {Chiang}}{2011}]{perez-becker_2011}
{Perez-Becker} D.,  {Chiang} E.,  2011, \apj, 735, 8

\bibitem[\protect\citeauthoryear{{Pilbratt}, {Riedinger}, {Passvogel}, {Crone},
  {Doyle}, {Gageur}, {Heras}, {Jewell}, {Metcalfe}, {Ott} \&
  {Schmidt}}{{Pilbratt} et~al.}{2010}]{pilbratt_2010}
{Pilbratt} G.~L.,  {Riedinger} J.~R.,  {Passvogel} T.,  {Crone} G.,  {Doyle}
  D.,  {Gageur} U.,  {Heras} A.~M.,  {Jewell} C.,  {Metcalfe} L.,  {Ott} S.,
  {Schmidt} M.,  2010, \aap, 518, L1

\bibitem[\protect\citeauthoryear{{Poglitsch}, {Waelkens}, {Geis} \& {et
  al}}{{Poglitsch} et~al.}{2010}]{poglitsch_2010}
{Poglitsch} A.,  {Waelkens} C.,  {Geis} N.,    {et al} 2010, \aap, 518, L2

\bibitem[\protect\citeauthoryear{{Rab}, {G{\"u}del}, {Padovani}, {Kamp}, {Thi},
  {Woitke} \& {Aresu}}{{Rab} et~al.}{2017}]{rab_2017}
{Rab} C.,  {G{\"u}del} M.,  {Padovani} M.,  {Kamp} I.,  {Thi} W.-F.,  {Woitke}
  P.,    {Aresu} G.,  2017, ArXiv e-prints

\bibitem[\protect\citeauthoryear{{Reames}}{{Reames}}{2013}]{reames_2013}
{Reames} D.~V.,  2013, \ssr, 175, 53

\bibitem[\protect\citeauthoryear{{Reedy}}{{Reedy}}{2012}]{reedy_2012}
{Reedy} R.~C.,  2012, in Lunar and Planetary Science Conference Vol.~43 of
  Lunar and Planetary Inst.~Technical Report, {Update on Solar-Proton Fluxes
  During the Last Five Solar Activity Cycles}.
p.~1285

\bibitem[\protect\citeauthoryear{{Salmeron} \& {Wardle}}{{Salmeron} \&
  {Wardle}}{2003}]{salmeron_2003}
{Salmeron} R.,  {Wardle} M.,  2003, \mnras, 345, 992

\bibitem[\protect\citeauthoryear{{Schlickeiser}}{{Schlickeiser}}{1989}]{schlickeiser_1989}
{Schlickeiser} R.,  1989, \apj, 336, 243

\bibitem[\protect\citeauthoryear{{Sicilia-Aguilar}, {Hartmann}, {Brice{\~n}o},
  {Muzerolle} \& {Calvet}}{{Sicilia-Aguilar}
  et~al.}{2004}]{sicilia-aguilar_2004}
{Sicilia-Aguilar} A.,  {Hartmann} L.~W.,  {Brice{\~n}o} C.,  {Muzerolle} J.,
  {Calvet} N.,  2004, \aj, 128, 805

\bibitem[\protect\citeauthoryear{{Turner} \& {Drake}}{{Turner} \&
  {Drake}}{2009}]{turner_2009}
{Turner} N.~J.,  {Drake} J.~F.,  2009, \apj, 703, 2152

\bibitem[\protect\citeauthoryear{{Umebayashi} \& {Nakano}}{{Umebayashi} \&
  {Nakano}}{1981}]{umebayashi_1981}
{Umebayashi} T.,  {Nakano} T.,  1981, \pasj, 33, 617

\bibitem[\protect\citeauthoryear{{Umebayashi} \& {Nakano}}{{Umebayashi} \&
  {Nakano}}{2009}]{umebayashi_2009}
{Umebayashi} T.,  {Nakano} T.,  2009, \apj, 690, 69

\bibitem[\protect\citeauthoryear{{Vacca} \& {Sandell}}{{Vacca} \&
  {Sandell}}{2011}]{vacca_2011}
{Vacca} W.~D.,  {Sandell} G.,  2011, \apj, 732, 8

\bibitem[\protect\citeauthoryear{{Wardle} \& {Salmeron}}{{Wardle} \&
  {Salmeron}}{2012}]{wardle_2012}
{Wardle} M.,  {Salmeron} R.,  2012, \mnras, 422, 2737

\bibitem[\protect\citeauthoryear{{Webb}, {Zuckerman}, {Platais}, {Patience},
  {White}, {Schwartz} \& {McCarthy}}{{Webb} et~al.}{1999}]{webb-1999}
{Webb} R.~A.,  {Zuckerman} B.,  {Platais} I.,  {Patience} J.,  {White} R.~J.,
  {Schwartz} M.~J.,    {McCarthy} C.,  1999, \apjl, 512, L63

\bibitem[\protect\citeauthoryear{{Webber}}{{Webber}}{1998}]{webber_1998}
{Webber} W.~R.,  1998, \apj, 506, 329

\bibitem[\protect\citeauthoryear{{Webber} \& {McDonald}}{{Webber} \&
  {McDonald}}{2013}]{webber_2013}
{Webber} W.~R.,  {McDonald} F.~B.,  2013, \grl, 40, 1665

\bibitem[\protect\citeauthoryear{{Woitke}, {Kamp} \& {Thi}}{{Woitke}
  et~al.}{2009}]{woitke_2009}
{Woitke} P.,  {Kamp} I.,    {Thi} W.-F.,  2009, \aap, 501, 383

\end{thebibliography}
